\documentclass[fleqn,usenatbib]{mnras}

\usepackage{newtxtext,newtxmath}

\usepackage[T1]{fontenc}

\DeclareRobustCommand{\VAN}[3]{#2}
\let\VANthebibliography\thebibliography
\def\thebibliography{\DeclareRobustCommand{\VAN}[3]{##3}\VANthebibliography}

\setcitestyle{notesep={ }}

\usepackage{graphicx}	%
\usepackage{amsmath}	%
\usepackage{multirow}
\usepackage{booktabs}

\title[Precision of Quasar IGM Damping Wing Measurements]{Quantifying the Precision of IGM Damping Wing Measurements Towards Quasars}

\author[Kist et al.]{
Timo Kist,$^{1}$\thanks{E-mail: kist@strw.leidenuniv.nl}
Joseph F. Hennawi$^{1,2}$
and Frederick B. Davies$^{3}$
\\
$^{1}$Leiden Observatory, Leiden University, P.O. Box 9513, 2300 RA Leiden,
The Netherlands\\
$^{2}$Department of Physics, University of California, Santa Barbara, CA 93106, USA\\
$^{3}$Max-Planck-Institut für Astronomie, Königstuhl 17, 69117 Heidelberg, Germany
}

\date{Accepted XXX. Received YYY; in original form ZZZ}

\pubyear{2024}

\begin{document}
\label{firstpage}
\pagerange{\pageref{firstpage}--\pageref{lastpage}}
\maketitle

\begin{abstract}
We investigate the precision with which the Lyman-$\alpha$ damping wing signature imprinted on the spectra of high-redshift quasars (QSOs) by the foreground neutral intergalactic medium (IGM) can measure the history of cosmic reionization. We leverage a novel inference pipeline based on a generative probabilistic model for the entire spectrum (both red- \textit{and} blueward of the Lyman-$\alpha$ line), accounting for all relevant sources of uncertainty -- the stochasticity caused by patchy reionization, the impact of the quasar's ionizing radiation on the IGM, it's unknown intrinsic spectrum, and spectral noise. Performing fast \texttt{JAX}-based Hamiltonian Monte-Carlo (HMC) parameter inference, we precisely measure the underlying global IGM neutral fraction as well as the lifetime of the quasar.
Running a battery of tests on over a thousand mocks, we find optimal precision when running the pipeline with a six parameter PCA continuum model (five coefficients and a normalization) on $\mathrm{S}/\mathrm{N} \sim 10$ spectra, binned to a $\sim 500\,\mathrm{km}/\mathrm{s}$ velocity pixel scale, and extending at least out to the \ion{C}{IV} $\lambda\,1549\,\text{\AA}$ emission line. After marginalizing out nuisance parameters associated with the quasar continuum, a single spectrum constrains the IGM neutral fraction to $28.0_{-8.8}^{\hspace{.056em}+\hspace{.056em}8.2}\,\%$ and the quasar lifetime to $0.80_{-0.55}^{\hspace{.056em}+\hspace{.056em}0.22}\,\mathrm{dex}$, improving notably towards spectra with a stronger IGM damping wing imprint. Higher precision can be achieved by averaging over statistical quasar samples. We identify two primary sources of uncertainty that  contribute approximately equally to the total error budget: the uncertain quasar continuum model and the stochastic distribution of neutral regions arising from both the reionization topology and the location of the quasar's ionization front.

\end{abstract}


\section{Introduction}

The Lyman-$\alpha$ damping wing signature of high-redshift quasars in the intergalactic medium (IGM) provides a unique way of probing the history of reionization. Being highly sensitive to the presence of intergalactic neutral hydrogen, Lyman-$\alpha$ absorption saturates at global volume-averaged IGM neutral fractions as small as $\langle x_\mathrm{HI} \rangle \sim 10^{-4}$ \citep{Gunn1965}. Even more importantly, at $\langle x_\mathrm{HI} \rangle = O(0.1)$, the absorption signature also starts to extend reward of the Lyman-$\alpha$ line in the form of the quantum-mechanical line-broadening that is well-described by the Lorentzian component of the Voigt profile \citep{Miralda-Escude1998}. This \textit{damping wing} absorption profile is a direct hallmark of the IGM optical depth and thus the reionization state of the IGM at the redshift of the observed object. %

Predicted by \citet{Miralda-Escude1998}, a quasar IGM
damping wing was first claimed for the $z = 7.09$ QSO ULAS J1120+0641 by \citet{Mortlock2011}. Re-analysis of the same object (as well as the $z = 7.54$ QSO ULAS J1342+0928) with different approaches has resulted in varying constraints on $\langle x_\mathrm{HI} \rangle$ \citep{Bolton2011, Greig2017b, Banados2018, Davies2018a, Greig2019, Durovcikova2020, Reiman2020}. To date, three distinct  analysis pipelines \citep{Greig2017a, Davies2018a, Durovcikova2020} have been applied to an additional number of QSOs at the highest redshifts currently accessible such as DES J0252-0503 at $z=7.00$ and J1007+2115 at $z=7.51$ \citep{Wang2020, Yang2020, Greig2022}. Lately, \citet{Durovcikova2024} and \citet{Greig2024b} presented the first analyses of larger QSO samples, probing the end stages of reionization between $5.8 \lesssim z \lesssim 7$. Moreover, a number of studies has recently suggested the presence of damping wing absorption adjacent to individual Gunn-Peterson troughs in the foreground of a quasar due to neutral islands in the $5.5 \lesssim z \lesssim 6$ IGM \citep{Becker2024, Zhu2024, Spina2024}. The number of available objects will increase drastically in the next years with the Euclid wide-field survey detecting $\gtrsim 100$ new quasars at $z>7$, and even objects out to $z\sim 9$ \citep[or higher, depending on the assumed quasar luminosity function (QLF);][]{Euclid2019}. Observations of their spectra 
will provide us with the first statistical QSO sample reaching deep into the heart of reionization.

The interest in damping wings has gained additional momentum recently with the first galaxy damping wings claimed in (stacks of) ultra high-redshift galaxy spectra observed by the James Webb Space Telescope (JWST) \citep{Curtis-Lake2023, Hsiao2023, Umeda2023, Keating2023a}. However, \citet{Heintz2023} pointed out a possibly significant contamination by intrinsic damped Lyman-$\alpha$ absorbers (DLAs) in some of these recently detected high-redshift galaxies, and \citet{Heintz2024} indeed identified strong integrated DLAs in a large fraction of galaxies ($60\,\%$ at $z \sim 6$ and up to $65 - 90\,\%$ at $z > 8$). Originating in the interstellar medium (ISM) or the circumgalactic medium (CGM), such proximate DLAs can interfere with or even dominate over the cosmologically relevant IGM damping wing imprint, potentially obstructing robust conclusions about the ionization state of the IGM. This underlines the importance of quasars as a complementary probe in the context of IGM damping wing analysis.

Though fewer in number as compared to galaxies, quasars come with the advantage that their strong ionizing radiation carves out a $\mathrm{Mpc}$-scale H II-region, the \textit{quasar proximity zone}, where all residual neutral CGM gas is ionized away, sidestepping the aforementioned problems related to intrinsic DLAs. Avoiding such issues with certainty, however, might require a strict exclusion of QSOs with proximate DLAs as recently pointed out in \citet{Davies2023} with regards to ULAS~J1342+0928. The task of identifying such objects will indeed be feasible for quasars (unlike in the context of much fainter objects such as galaxies) by using the capabilities of JWST or ELT to identify weak metal absorption systems in their spectra \citep{Davies2023}.

An additional virtue of quasars as compared to galaxies is the simplicity in modelling their continuum that the IGM damping wing is imprinted upon. The reconstruction of this continuum acts as a nuisance stochastic process when constraining $\langle x_\mathrm{HI} \rangle$ based on the IGM damping wing imprint. Owing to the absence of evolution between $2 \lesssim z \lesssim 5$ as evidenced by \citet{Shen2007}, it stands to reason to construct empirical quasar continuum models based on large datasets of $O(1000-10\,000)$ low-redshift objects from surveys such as SDSS \citep[see][for a comprehensive overview and comparison of methods]{Greig2024a}.

The strength of the IGM damping wing imprint is (to leading order) not only affected by the global volume-averaged IGM neutral fraction $\langle x_\mathrm{HI} \rangle$ at the quasar's redshift but also by its light curve: the longer a quasar has been shining, the larger the ionized region it has carved out in its vicinity, manifesting in a weaker damping wing imprint and a more extended proximity zone. Under the assumption of a light-bulb light curve, this is determined by a single nuisance parameter, the lifetime of the quasar. While acting as a nuisance parameter when constraining the history of reionization, the quasar lifetime also bears physical information on its own about the formation and growth of supermassive black holes (SMBHs) whose understanding is increasingly challenged by JWST observations of moderately massive SMBHs at ultra-high redshifts \citep{Uebler2024, Maiolino2023, Kokorev2023, Larson2023, Bogdan2024}. Proximity zone sizes as simple summary statistics informing about the lifetime of a quasar have been measured extensively in the literature \citep{Eilers2017, Eilers2020, Ishimoto2020, Satyavolu2023b}, and remain the subject of ongoing modelling efforts \citep[e.g.][]{Davies2020, Satyavolu2023a, Zhou2024}.

Further complications arise due to the fact that reionization is a patchy process: the local size of reionization bubbles at a given value of $\langle x_\mathrm{HI} \rangle$ can vary significantly \citep{Davies2018a}, especially during the mid-stages of reionization, hence acting as an additional source of stochasticity when inferring $\langle x_\mathrm{HI} \rangle$. Besides that, fluctuations in the density field source stochasticity in the size of the ionized bubble carved out by the quasar \citep{Davies2018a}. These effects can only be accounted for with cosmological hydrodynamical simulations including ionizing radiative transfer.

\citet{hennawi2024} introduced a fully Bayesian approach 
to jointly infer the global volume-averaged IGM neutral fraction $\langle x_\mathrm{HI} \rangle$, quasar lifetime $t_\mathrm{Q}$ and the underlying continuum of an observed quasar spectrum, accounting for the full covariance arising from continuum reconstruction as well as the Lyman-$\alpha$ transmission stochastic process. The pipeline entails a parametric model for the quasar continuum based on principal component analysis (PCA), constructed from a large sample of $\sim 15\,000$ low-redshift ($2.149 < z < 4$) continua. IGM transmission fields describing the IGM damping wing are constructed by combining skewers from the hydrodynamical Nyx simulations \citep{Almgren2013, Lukic2015} and semi-numerical reionization topologies \citep{Mesinger2011, Davies2022} via 1d radiative transfer \citep{Davies2016}. \citet{hennawi2024} introduced a likelihood prescription that allows us to \textit{jointly} fit the continuum and the IGM damping wing imprint, accounting for all uncertainties due to the stochasticity of the reionization topology, the unknown qusar lifetime, continuum reconstruction errors, and spectral noise. In addition, the likelihood operates on the entire spectrum and hence overcomes the limitations of the red-blue split commonly performed in previous approaches \citep[e.g.][]{Greig2017a, Davies2018a, Durovcikova2020, Reiman2020} which by construction cannot use the information on the blue side of the spectrum to recover the continuum.

A rigorous quantification of the associated error budget is key to deriving competitive cosmological and astrophysical constraints on $\langle x_\mathrm{HI} \rangle$ and $t_\mathrm{Q}$ from observed high-redshift quasar spectra. The scope of our work in this context is twofold: 1) we determine the optimal observational setup to obtain maximally tight yet statistically faithful constraints, and 2) we break down the error budget across parameter space into the individual contributions of all model components.

Inference precision will inevitably be affected by hyperparameters of the underlying models such as the dimension and hence the flexibility of our parametric continuum model. Additional impact comes from the specifics of the observational setup
such as the covered wavelength range, signal-to-noise ratio or spectral resolution. The inference performance with respect to these parameters informs us about the optimal observational strategy
and the cost-benefit ratio of modern-day instruments for the purposes of our analysis. Observational considerations aside, the aforementioned parameters can also have a profound impact on the computational complexity of the inference problem at hand. Given the cubic scaling of the evaluation of Gaussian likelihood terms with the number of spectral pixels, a too ambitious choice of the aforementioned hyper-parameters can easily make the problem computationally expensive or even intractable.
This is particularly true for the highest spectral resolutions that are already achievable with modern-day instruments. Our analysis explores these dependencies and makes practical suggestions for the optimal parameter setting for the inference machinery introduced in \citet{hennawi2024}. 

Secondly, inference precision intrinsically depends on the values of the physical parameters that are being inferred. A lower IGM neutral fraction $\langle x_\mathrm{HI} \rangle$, for example, or a longer quasar lifetime $t_\mathrm{Q}$ lead to a weaker damping wing imprint, making the parameter values harder to infer. Additional uncertainty arises from the two major stochastic processes that are part of our pipeline: 
the parametric quasar continuum model and the Lyman-$\alpha$ transmission model accounting for the patchiness of reionization and stochasticity in the position of the quasar ionization front due to density fluctuations.
We rigorously quantify their contributions to the overall uncertainty on the inferred IGM neutral fraction $\langle x_\mathrm{HI} \rangle$ and quasar lifetime $t_\mathrm{Q}$, and show how these uncertainties vary as a function of astrophysical parameter space.

Section~\ref{sec:methods} provides a short but largely self-contained summary of the inference pipeline introduced in \citet{hennawi2024}. We present our hyper-parameter sensitivity analysis in Section~\ref{sec:sensitivity}, culminating in the full breakdown of the total error budget in Section~\ref{sec:error_budget}. We conclude in Section~\ref{sec:conclusions}. In Appendix~\ref{app:mortlock} we correct a typo in the analytical damping wing optical depth equation provided by \citet{Mortlock2016}. Appendix~\ref{app:inf_tests} summarizes the inference tests we performed to demonstrate the statistical fidelity of our pipeline, and Appendix~\ref{app:param_scan_sims} extends the sensitivity analysis presented in Section~\ref{sec:param_scan} to our full forward model based on cosmological simulations.

\section{Methods}
\label{sec:methods}

We employ the inference procedure 
introduced in \citet{hennawi2024} to infer the global volume-averaged neutral fraction $\langle x_\mathrm{HI} \rangle$ and quasar lifetime $t_\mathrm{Q}$ with help of the damping wing signature that appears in high-redshift quasar spectra. We will restrict this section to a short summary of the formalism, and refer the reader to \citet{hennawi2024} for additional details.

The observable in this context is the spectrum $\boldsymbol{f}$ of the quasar that comes with a noise vector $\boldsymbol{\sigma}$. Inferring the above mentioned two astrophysical parameters $\langle x_\mathrm{HI} \rangle$ and $t_\mathrm{Q}$ given a measurement of $\boldsymbol{f}$ and $\boldsymbol{\sigma}$ requires a forward model of IGM damping wing absorption and an expression for the likelihood of the observed spectrum given the parameters of our model. To that end, we assume that the full observed flux $\boldsymbol{f}$ arises from a quasar continuum $\boldsymbol{s}$ that is subject to an IGM transmission field $\boldsymbol{t}$ and spectral noise $\boldsymbol{\sigma}$. 
In practice, we do not have perfect knowledge of $\boldsymbol{s}$ and $\boldsymbol{t}$ as we cannot observe each of them separately. In other words, they constitute \textit{latent variables} of our model. To move forward, we assume that they are governed by stochastic processes: we introduce a new low-dimensional latent variable $\boldsymbol{\eta}$ that represents the full quasar continuum $\boldsymbol{s}$, and we generate IGM transmission fields $\boldsymbol{t}$ based on simulations that take as input a set of astrophysical parameters $\boldsymbol{\theta}$. \citet{hennawi2024} showed that under the assumption of Gaussianity for the stochastic processes governing $\boldsymbol{s}$ and $\boldsymbol{t}$, the full likelihood of measuring the flux $\boldsymbol{f}$ can be 
approximated as
\begin{equation}
\label{eq:likelihood}
    L(\boldsymbol{f}|\boldsymbol{\sigma}, \boldsymbol{\theta}, \boldsymbol{\eta}) = \mathcal{N}\left(\boldsymbol{f} ;\langle\boldsymbol{t}\rangle \circ\langle\boldsymbol{s}\rangle, \boldsymbol{\Sigma}+\langle\boldsymbol{S}\rangle \boldsymbol{C}_{\boldsymbol{t}}\langle\boldsymbol{S}\rangle+\langle\boldsymbol{T}\rangle \boldsymbol{C}_{\boldsymbol{s}}\langle\boldsymbol{T}\rangle\right),
\end{equation}
where $\mathcal{N}(\boldsymbol{f}; \boldsymbol{\mu}, \boldsymbol{K})$ is the standard normal distribution with mean $\boldsymbol{\mu}$ and covariance matrix $\boldsymbol{K}$. Throughout, $\boldsymbol{x}\circ\boldsymbol{y}$ denotes the element-wise (Hadamard) product of two vectors $\boldsymbol{x}$ and $\boldsymbol{y}$, and bold lowercase symbols denote vectors, bold uppercase symbols matrices. Here we defined $\boldsymbol{\Sigma} \equiv \mathrm{diag}(\boldsymbol{\sigma})$,  $\boldsymbol{S}\equiv\mathrm{diag}(\boldsymbol{s})$, and $\boldsymbol{T}\equiv\mathrm{diag}(\boldsymbol{t})$, with $\boldsymbol{C}_{\boldsymbol{s}}$ and $\boldsymbol{C}_{\boldsymbol{t}}$ as the corresponding covariance matrices.

We emphasize that a red-blue split of the spectrum is not required at any stage of the derivation of Eq.~(\ref{eq:likelihood}). Our forward model is a generative process for the full flux across the entire spectral range, and hence Eq.~(\ref{eq:likelihood}) also takes into account information from the blue side of the spectrum, most particularly the region of the smooth IGM damping wing signature and the quasar proximity zone. This implies that we are not restricted to the commonly adopted approach of predicting the blue-side continuum \textit{before} fitting for the optimal parameters $\boldsymbol{\theta}$ on the resulting continuum-normalized spectrum \citep{Greig2017a, Greig2017b, Davies2018a, Davies2018b, Banados2018, Greig2019, Durovcikova2020, Wang2020, Yang2020}. Instead, we infer the astrophysical parameters $\boldsymbol{\theta}$ \textit{jointly} with the nuisance parameters $\boldsymbol{\eta}$ representing the continuum across the entire spectral range.

The approach of considering the full spectrum only gained momentum very recently in the context of low-redshift quasar continuum prediction \citep{Sun2023}, and has been introduced for high-redshift quasar damping wings in \citet{hennawi2024}. It comes at the price of errors and bias resulting from the fact that Eq.~(\ref{eq:likelihood}) is approximate
due to
the non-Gaussianity of the IGM transmission stochastic process \citep{Lee2015, Davies2018a}. The approximate nature of our likelihood prescription can result in misplaced or too narrow posterior distributions, and we correct for the concomitant loss of information when determining inference precision. This paper 
rigorously quantifies the errors resulting down the line in the inference process, and explores the optimal parameter setting to maximize inference precision with a view on observational and computational cost of the inference task at hand.

\subsection{Training Data and Dimensionality Reduction}
\label{sec:dim_red}

Lacking a principled physical model for quasar continua, we adopt a data-driven approach to construct a lower-dimensional parametric model $\boldsymbol{s}_\mathrm{DR}(\boldsymbol{\eta})$ of the full continuum $\boldsymbol{s}$. Due to the strong correlations among individual pixels, it is not favorable nor necessary to infer the value of $\boldsymbol{s}$ at every single spectral pixel. Instead, we make use of these inter-pixel correlations to reduce the dimensionality of the continuum from the total number of spectral pixels $n_\lambda$ to a lower-dimensional latent representation, fully specified by the new $n_\mathrm{latent}$-dimensional latent variable $\boldsymbol{\xi}$ (with $n_\mathrm{latent} \ll n_\lambda$). A simple and well-established approach to do so is based on principal component analysis (PCA) \citep[e.g.][]{Suzuki2005, Suzuki2006, Paris2011, Davies2018b}, resulting in the expression
\begin{equation}
\label{eq:pca}
    \boldsymbol{s}_\mathrm{DR}(\boldsymbol{\xi}) = \langle \boldsymbol{s} \rangle + \boldsymbol{\xi}^T\boldsymbol{A}.
\end{equation}
Here $\langle \boldsymbol{s} \rangle$ denotes the mean of the $n_\mathrm{train}$ training continua $\boldsymbol{s}$; $\boldsymbol{A}$ is the $(n_\mathrm{latent}\times n_\lambda)$-dimensional matrix of PCA basis vectors, and $\boldsymbol{\xi}$ the $n_\mathrm{latent}$-dimensional vector of PCA coefficients. Along with the amplitude $s_\mathrm{norm}$ of the continuum at $\lambda = 1450\,\text{\AA}$, we end up with a $(n_\mathrm{latent}+1)$-dimensional parametric model $\boldsymbol{s}_\mathrm{DR}(\boldsymbol{\eta})$ for the quasar continuum, where $\boldsymbol{\eta} = (s_\mathrm{norm}, \boldsymbol{\xi})$. Again, we stress that our model represents the continuum across the entire spectral range without a division into a blue-side and a red-side part.

Our choice of a standard PCA decomposition as dimensionality reduction (DR) method is motivated by the fact that \citet{hennawi2024} showed it to perform comparably well to more complex machine learning-based techniques using Gaussian processes or variational autoencoders. To construct the PCA model, we assume that we can always find an (unabsorbed) low-redshift quasar spectrum that is a close representative of a given high-redshift continuum.\footnote{Note that we are \textit{not} assuming that intrinsic shape of quasar continua does not change as a function of redshift. It suffices to make the weaker assumption that we see examples of high-redshift quasars at low $z$. The opposite, i.e., that the full plethora of shapes we observe at low redshift also has to be present at high $z$, does not necassarily have to be true.} 
More specifically, we make use of $15\,559$ low-redshift quasar spectra from the SDSS-III Baryon Oscillation Spectroscopic Survey (BOSS) and SDSS-IV Extended BOSS (eBOSS) with SDSS autofits. The resolution of the spectra is $R \sim 2000$ and their redshifts are in the range $2.149 < z < 4$. The spectra are chosen such that each of them covers the full rest-frame wavelength range between $1170\,\text{\AA}$ and $2040\,\text{\AA}$ and that their median signal-to-noise ratio is $\mathrm{S}/\mathrm{N} > 10$ within a $5.0\,\text{\AA}$ region centered around the rest-frame wavelength $1285\,\text{\AA}$. In order to allow for a joint PCA decomposition of all training spectra, we are rebinning the spectra onto a common rest-frame wavelength grid linearly spaced in velocity with pixels of size $\mathrm{d}v_\mathrm{pix} = 140\;\mathrm{km}/\mathrm{s}$. For further details, we refer the reader to \citet{hennawi2024}. We divide the set of continua into $n_\mathrm{train} = 14\,781$ training objects to determine the PCA basis $\boldsymbol{A}$ and $n_\mathrm{test} = 778$ test objects to draw mock continua from and estimate the continuum reconstruction error. This corresponds to a $95\,\%-5\,\%$ training-test split. We account for the reconstruction error stochastic process by defining the 
relative continuum reconstruction error as
\begin{equation}
\label{eq:reconstr_err}
    \boldsymbol{\delta} = \frac{\boldsymbol{s}-\boldsymbol{s}_\mathrm{DR}(\boldsymbol{\eta})}{\boldsymbol{s}},
\end{equation}
where the division by $\boldsymbol{s}$ is to be understood as an element-wise operation. As shown in \citet{hennawi2024}, $\boldsymbol{\delta}$ closely follows a Gaussian distribution $\mathcal{N}(\boldsymbol{\delta}; \langle\boldsymbol{\delta}\rangle, \Delta)$, and we estimate the mean $\langle\boldsymbol{\delta}\rangle$ and the covariance matrix $\Delta$ by applying our DR formalism to the $n_\mathrm{test}=778$ test objects. Eq.~(\ref{eq:reconstr_err}) implies that $\boldsymbol{s} \simeq \boldsymbol{s}_\mathrm{DR}(\boldsymbol{\eta})\circ(\boldsymbol{1}+\boldsymbol{\delta})$ up to linear order in $\boldsymbol{\delta}$, and therefore we obtain
\begin{equation}
\label{eq:P_s_alpha}
    P(\boldsymbol{s}|\boldsymbol{\eta}) = \mathcal{N}(\boldsymbol{s}; \langle\boldsymbol{s}(\boldsymbol{\eta})\rangle, \boldsymbol{C}_{\boldsymbol{s}}(\boldsymbol{\eta}))
\end{equation}
if we define $\langle\boldsymbol{s}(\boldsymbol{\eta})\rangle \equiv \boldsymbol{s}_\mathrm{DR}(\boldsymbol{\eta})\circ(\boldsymbol{1}+\langle\boldsymbol{\delta}\rangle)$,  and $\boldsymbol{C}_{\boldsymbol{s}}(\boldsymbol{\eta}) \equiv \mathrm{diag}(\boldsymbol{s}_\mathrm{DR}(\boldsymbol{\eta}))\;\Delta\;\mathrm{diag}(\boldsymbol{s}_\mathrm{DR}(\boldsymbol{\eta}))$. Note that $\langle\boldsymbol{\delta}\rangle$ and $\Delta$ (and hence also $\langle\boldsymbol{s}(\boldsymbol{\eta})\rangle$ and $\boldsymbol{C}_{\boldsymbol{s}}(\boldsymbol{\eta})$) depend on the choice of latent dimension $n_\mathrm{latent}$ of the DR model, and are thus recomputed whenever we vary $n_\mathrm{latent}$.

\subsection{Simulating Damping Wing Observations}
\label{sec:simulators}

To model IGM damping wing absorption profiles, we generate IGM transmission fields $\boldsymbol{t}$, considering their variation with respect to two astrophysical parameters $\boldsymbol{\theta} = (\langle x_\mathrm{HI} \rangle, t_\mathrm{Q})$, where $\langle x_\mathrm{HI} \rangle$ is the global volume-averaged IGM neutral hydrogen fraction, and $t_\mathrm{Q}$ is the lifetime of the quasar. We explore two different approaches of generating $\boldsymbol{t}$ for a given set of parameters $\boldsymbol{\theta}$:
1) Following \citet{Davies2018a}, we post-process sightlines from the cosmological hydrodynamical Nyx simulations \citep{Almgren2013, Lukic2015} and \texttt{21cmFAST} reionization topologies \citep{Mesinger2011} with 1d ionizing radiative transfer \citep{Davies2016}, and 2), for the purpose of breaking up the error budget into the separate contributions of all individual model components, we eliminate the stochasticity between $\boldsymbol{\theta}$ and the simulated IGM damping wing strength by constructing a simple analytical model that relates them deterministically. We introduce these two approaches in the following sections.

\subsubsection{Full-Simulation Model}
\label{sec:sims}

To obtain maximally realistic IGM transmission fields, we follow the procedure introduced in \citet{Davies2018a}: we extract $1200$ density, velocity, and temperature skewers from a snapshot of the Nyx hydrodynamical simulations \citep{Almgren2013, Lukic2015} at redshift $z=7.0$. The simulations have a box size of $100\;\mathrm{cMpc}/h$, and contain $4096^3$ baryon and dark matter particles each. A realistic reionization topology is generated with help of a modified version \citep{Davies2022} of the \texttt{21cmFast} code \citep{Mesinger2011}, tuning the ionizing efficiency $\zeta$ to obtain $21$ different ionization fields with global volume-averaged neutral fractions between $\langle x_\mathrm{HI} \rangle = 0$ and $\langle x_\mathrm{HI} \rangle = 1$ in steps of  $\Delta \langle x_\mathrm{HI} \rangle = 0.05$. From each run, we extract $10 000$ randomly oriented $x_\mathrm{HI}$ skewers originating at the $500$ most massive halos. We combine each of the $1200$ Nyx density and temperature skewers with a random $x_\mathrm{HI}$ skewer by performing one-dimensional radiative transfer along the sightlines \citep{Davies2016}, integrating up to $51$ different quasar lifetimes between $t_\mathrm{Q} = 10^3\,\mathrm{yr}$ and $t_\mathrm{Q} = 10^8\,\mathrm{yr}$ in steps of $\Delta\log_{10}(t_\mathrm{Q}/\mathrm{yr}) = 0.1$ \citep{Davies2019}. For this, we assume a flat $\Lambda$CDM cosmology with $h=0.685$, $\Omega_b = 0.047$, $\Omega_m = 0.3$, $\Omega_\Lambda = 0.7$,  and $\sigma_8 = 0.8$.
After excluding $17$ of the resulting skewers because of strong proximate DLA absorption, this yields a $21\times 51$ grid in $\boldsymbol{\theta} = (\langle x_\mathrm{HI} \rangle, t_\mathrm{Q})$ with $1183$ distinct Lyman-$\alpha$ transmission fields at each point in parameter space.

Generally, the IGM transmission stochastic process $P(\boldsymbol{t}|\boldsymbol{\theta})$ is well known to be a non-Gaussian \citep{Lee2015, Davies2018a}. Nevertheless, for the sake of deriving an analytical expression (Eq.~(\ref{eq:likelihood})) for the full likelihood $L(\boldsymbol{f}|\boldsymbol{\sigma}, \boldsymbol{\theta}, \boldsymbol{\eta})$, we approximated it as such, i.e.,
\begin{equation}
\label{eq:P_t_theta}
    P(\boldsymbol{t}|\boldsymbol{\theta}) = \mathcal{N}(\boldsymbol{t}; \langle\boldsymbol{t}(\boldsymbol{\theta})\rangle, \boldsymbol{C}_{\boldsymbol{t}}(\boldsymbol{\theta})).
\end{equation}
In this context, we estimate the mean $\langle\boldsymbol{t}(\boldsymbol{\theta})\rangle$ and the covariance $\boldsymbol{C}_{\boldsymbol{t}}(\boldsymbol{\theta})$ empirically from the aforementioned simulated realizations. As shown in \citet{hennawi2024}, this is a reasonably close approximation at sufficiently low spectral resolution. The tests in Appendix~\ref{app:inf_tests} reveal that this is not the case at higher resolutions where Eq.~(\ref{eq:P_t_theta}) can lead to notable biases in our parameter inference that we have to take into account when evaluating inference precision. We describe a structured way to do so in Section~\ref{sec:inf_tests}.

\subsubsection{Analytical Simulator}
\label{sec:analytic}

\begin{figure}
    \centering
	\includegraphics[width=\columnwidth]{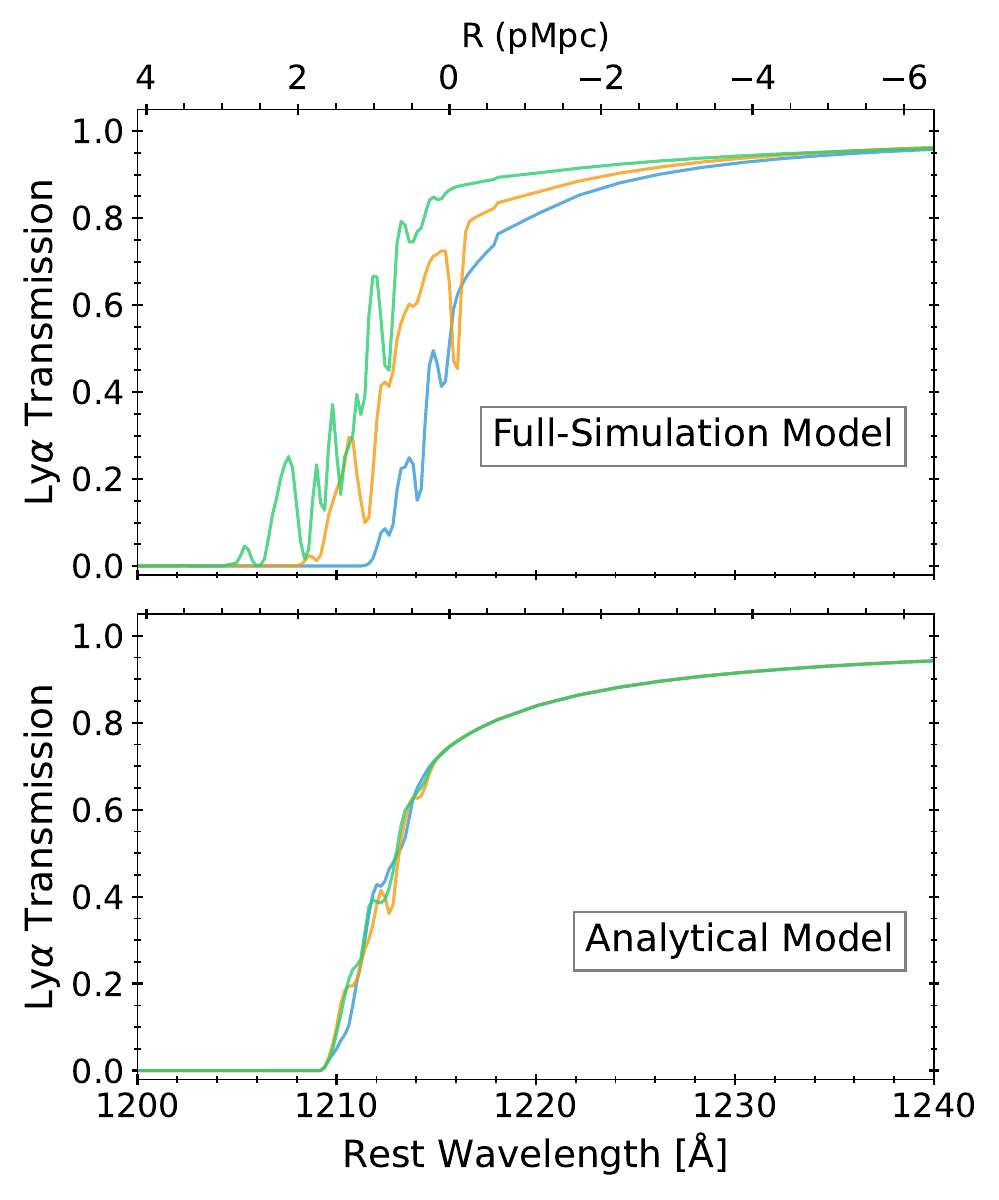}
    \caption{Three random examples of IGM transmission skewers generated with the full-simulation model (see Section~\ref{sec:sims}) and the analytical model (see Section~\ref{sec:analytic}). All skewers were generated assuming a global IGM neutral fraction of $\langle x_\mathrm{HI} \rangle = 0.5$ and a quasar lifetime of $t_\mathrm{Q} = 10^6\,\mathrm{yr}$. The size of the ionization front varies significantly between the three full-simulation skewers due to the stochasticity of the ionized bubble size distribution that is incorporated in this model but not in the analytical one.}
    \label{fig:skewers}
\end{figure}

While the aforementioned numerical simulator captures the relevant
physics for modeling quasar proximity zones and IGM damping wings, one
potential drawback is that the size of the proximity zone and strength
of the IGM damping wing are stochastically related to the two
astrophysical parameters $\langle x_{\rm HI}\rangle$ and
$\log_{10}(t_{\rm Q}/{\rm yr})$. For example, for a fixed global volume averaged
neutral fraction $\langle x_{\rm HI}\rangle$, the reionization
topology results in a broad distribution of distances to the nearest
patch of neutral hydrogen \citep[see Figure~2
  of][]{Davies2018a}. In addition, small-scale overdensities
corresponding to Lyman limit systems or partial Lyman
limit sytems (LLS or PLLS) can significantly attenuate the quasar radiation, and the
presence or absence of such fluctuations produce scatter in the
location of the quasar ionization front \citep{Chen2021}. These physical effects, which ultimately arise from
density fluctuations on a hierarchy of scales, source stochasticity in
the proximity zone size and the strength of the IGM damping wing.

It is conceivable that an alternative parameterization exists that
would allow one to fit for the strength of the IGM damping wing as set
by the total column density of neutral gas in the foreground of the
quasar, which could potentially remove much of the stochasticity between
damping wing strength and the parameters $\langle x_{\rm HI}\rangle$
and $\log_{10}t_{\rm Q}$. From the standpoint of the goals of this
paper, a more deterministic relationship between damping wing
strength and model parameters would be very valuable, as it would
allow us to isolate how factors like continuum reconstruction errors, latent space dimension, ${\rm S\slash N}$ ratio, and spectral
resolution contribute to the overall error budget; whereas, the
relative contributions of these factors will be concealed for the 
more stochastic numerical simulator described in the previous section, once the intrinsic scatter in
damping wing strength at fixed $\btheta = \{\langle x_{\rm HI}\rangle,
\log_{10}t_{\rm Q}\}$ dominates the overall noise budget. To this end, we also
explore a more deterministic analytical IGM damping wing simulator, which we now describe.

For simplicity, to generate the IGM damping wing we consider an isotropically emitting quasar in an
IGM of uniform density and uniform neutral fraction. In a second step, we then allow for density fluctuations in the proximity zone via a lognormal Lyman-$\alpha$ forest model.

We start by writing the mean neutral hydrogen number density at the redshift $z_\mathrm{QSO}$ of the quasar as
\begin{equation}
    \langle n_{\rm HI} \rangle = n_{\rm H, 0} (1+z_\mathrm{QSO})^3 \langle x_{\rm HI}\rangle,
\end{equation}
where $n_{\rm H, 0}$ is the present mean hydrogen number density and $\langle x_{\rm HI}\rangle$ is the global volume-averaged IGM neutral fraction at $z_\mathrm{QSO}$. Here $n_{\rm H, 0} = 3H_0^2(1-Y)\Omega_{\rm b}/8\pi G m_{\rm p} \simeq 1.9\times10^{-7} \; {\rm atoms}/{\rm cm}^3$ 
with big bang nucleosynthesis (BBN) hydrogen fraction $1-Y = 0.76$. All cosmological parameters for the analytical simulator described in this section are set according to a \citet{Planck2020} cosmology. 
Assuming that each photon emitted by the quasar ionizes precisely one hydrogen atom in the quasar's immediate vicinity and ignoring recombinations as well as Hubble expansion, the radius $R_{\rm ion}$ of the resulting ionization front evolves with the lifetime $t_\mathrm{Q}$ of the quasar as
\begin{eqnarray}
\label{eq:R_ion}
R_{\rm ion} &=& \left(\frac{3 Q t_{\rm Q}}{4\pi \langle n_{\rm HI}\rangle }\right)^{1\slash 3}\\
          &=& 12.0\,{\rm cMpc}\left(\frac{Q}{10^{57.1}\,{\rm s^{-1}}}\right)^{1\slash 3}\left(\frac{t_{\rm Q}}{10^6\,{\rm yr}}\right)^{1\slash 3}\left(\frac{\langle x_{\rm HI}\rangle}{1.0}\right)^{-1\slash 3}\nonumber
\end{eqnarray}
\citep{Cen2000}, 
where $Q$ is the quasar's emission rate of ionizing photons, obtained by integrating the luminosity $L_\nu$ per photon energy $h\nu$ over all frequencies larger than the Lyman-limit frequency $\nu_{\rm LL}$:
\begin{equation}
Q = \int_{\nu_{\rm LL}}^{\infty} \frac{L_{\nu}}{h\nu} d\nu.
\end{equation}
Here we assume that the luminosity follows a power-law shape $L_\mathrm{\nu} = L_\mathrm{LL}\,(\nu/\nu_\mathrm{LL})^{-\alpha_\mathrm{s}}$ blueward of the Lyman limit frequency $\nu_\mathrm{LL}$ with a spectral index $\alpha_\mathrm{s} = 1.7$ and $L_\mathrm{LL}$ set by the $J$-band magnitude and redshift of the quasar which we choose as $z_\mathrm{QSO}=7.54$ and $J=20.3$, resembling the quasar ULAS J1342+0928. Altogether, this amounts to $Q = 10^{57.1}\,{\rm s^{-1}}$.
\\

The IGM transmission field $t(\lambda_{\rm obs})$ as a function of observed-frame wavelength $\lambda_{\rm obs} = (1+z_\mathrm{QSO})\,\lambda_{\rm rest}$ is governed by the Lyman-$\alpha$ optical depth $\tau(\lambda_{\rm obs})$ as
\begin{equation}
    t(\lambda_{\rm obs}) = e^{-\tau(\lambda_{\rm obs})}.
\end{equation}
We obtain the optical depth $\tau(\lambda_{\rm obs})$ by integrating the infinitesimal optical depth $\mathrm{d}\tau = \langle n_\mathrm{HI}\rangle \,\sigma_\alpha(\nu)\;\mathrm{d}l$ along a line-of-sight interval $\mathrm{d}l$
\citep{Miralda-Escude1998, Mortlock2016}:
\begin{equation}
\label{eq:tau_int}
    \tau_{\rm DW}(\lambda_{\rm obs}) = \int \langle n_\mathrm{HI}\rangle \,\sigma_\alpha(\nu)\;\mathrm{d}l.
\end{equation}
Here $\sigma_\alpha(\nu)$ is the scattering cross section of the Lyman-$\alpha$ line which is given as
\begin{equation}
\label{eq:cross_section}
    \sigma_\alpha(\nu) = \frac{\pi e^2}{m_e c}\; f_\alpha \; \phi_\alpha(\nu),
\end{equation}
where $f_\alpha\simeq 0.416$ is the Lyman-$\alpha$ oscillator strength and $\phi_\alpha(\nu)$ the line profile function. We assume that $\phi_\alpha(\nu)$ is of Lorentzian shape
\begin{equation}
\label{eq:line_profile}
    \phi_\alpha(\nu) = \frac{R_\alpha/(\pi\nu_\alpha)}{(\nu/\nu_\alpha-1)^2+R_\alpha^2}
\end{equation}
with Lyman-$\alpha$ frequency $\nu_\alpha = c/\lambda_\alpha$, Lyman-$\alpha$ wavelength $\lambda_\alpha = 1215.67\,\text{\AA}$, and $R_\alpha = \Gamma_\alpha/4\pi\nu_\alpha$ determined by the decay constant $\Gamma_\alpha=6.265\,\times\,10^8\,\mathrm{s}^{-1}$ of the Lyman-$\alpha$ transition. \citet{Bach2015} showed this to be a more accurate approximation to the full quantum-mechanical cross section than the commonly used two-level model by \citet{Miralda-Escude1998}. Integrating Eq.~(\ref{eq:tau_int}) from the redshift $z_{\rm ion}$ of the ionization front to the end of reionization (that we assume
at $z_{\rm end} = 6.0$; our results are not sensitive to this choice)
yields the damping wing optical depth
\begin{eqnarray}
\label{eq:tau_DW}
  \tau_{\rm DW}(\lambda_{\rm obs}) &=& \frac{4 R_\alpha \tau_{\rm GP}(z_{\rm ion})}{\pi}\left(\frac{\lambda_{\rm obs}}{(1 + z_{\rm ion})\lambda_\alpha}\right)^{3\slash 2}\times\\
  &\!&I\left[\frac{(1 + z_{\rm ion})\lambda_\alpha}{\lambda_{\rm obs}}\right] -   I\left[\frac{(1 + z_{\rm end})\lambda_\alpha}{\lambda_{\rm obs}}\right]\nonumber.
\end{eqnarray}
Here $\tau_{\rm GP}(z)$ is the Gunn-Peterson optical depth
\begin{equation}
\label{eq:tau_GP}
\tau_{\rm GP}(z) = 5.3\times 10^5 \left(\frac{1+z}{1+7.54}\right)^\frac{3}{2} \left(\frac{H_0\sqrt{\Omega_\mathrm{m}(1+z)^3}}{H(z)}\right) \left(\frac{\langle x_{\rm HI}\rangle}{1.0}\right),
\end{equation}
and $I(x)$ is a dimensionless integral factor given by
\begin{equation}
\label{eq:integral}
I(x) = \frac{x^{1\slash 2}}{4(1 - x)} +\frac{1}{8} \ln\left(\frac{1- x^{1\slash 2}}{1 + x^{1\slash 2}}\right),
\end{equation}
following the derivation in \citet{Mortlock2016} 
\footnote{Note that due to a typo in \citet{Mortlock2016}, the prefactor of the second term in Eq.~(\ref{eq:integral}) differs with respect to 
that in Eq.~(40) in \citet{Mortlock2016}. We elaborate on the details and consequences of this correction in Appendix~\ref{app:mortlock}.}.

Finally, to model the Lyman-$\alpha$ forest fluctuations in the quasar proximity zone, we start by generating realizations of a lognormal model using a modified version of the approach adopted by \citet{Karacayli2020}, which is itself a modification of \citet{McDonald2006}. In this approach, the non-linear and non-Gaussian HI column density field (and as a result, the optical depth field) is generated by applying a squared lognormal transformation to a Gaussian baryon field, reproducing the theoretically expected mean flux ${\bar F}$ evolution and power spectrum of the Lyman-$\alpha$ forest. We draw realizations of the lognormal model at a reference redshift of $z_\mathrm{ref} = 4.95$ and scale them to the quasar redshift $z_\mathrm{QSO} = 7.54$ assuming a metagalactic UVB photoionization rate of $\Gamma_{\rm HI} = 10^{-13}\;1/\mathrm{s}$ at that redshift. Combining these high-redshift Lyman-$\alpha$ forest profiles with the aforementioned  damping wing profiles provides us with a full Lyman-$\alpha$ transmission model that we can compare to the (more realistic) full-simulation model introduced in Section~\ref{sec:sims}.

We show three random skewers produced by each of the two models in Figure~\ref{fig:skewers}. All skewers were generated at a fixed IGM neutral fraction of $\langle x_\mathrm{HI} \rangle = 0.5$ and a fixed quasar lifetime of $t_\mathrm{Q} = 10^6\,\mathrm{yr}$, yet the three skewers based on the full-simulation model show significant differences. In particular, the sizes of their proximity zones vary between $\sim 1.0 - 2.5\;\mathrm{pMpc}$, 
reflective of the stochastic bubble size distribution in the underlying semi-numerical reionization topology. By construction, the three analytical skewers look almost identical as they only differ by the draw of the lognormal Lyman-$\alpha$ forest fluctuations. This illustrates how the analytic formalism introduced in this section eliminates the stochasticity of reionization from our forward modelling, allowing us to measure the impact of other sources of uncertainty more accurately, and quantify the extent as to which the intrinsic stochasticity of reionization contributes to the total error budget.

\subsection{Generating Mock Spectra}
\label{sec:mock}

\begin{figure*}[h]
    \centering
	\includegraphics[width=\textwidth]{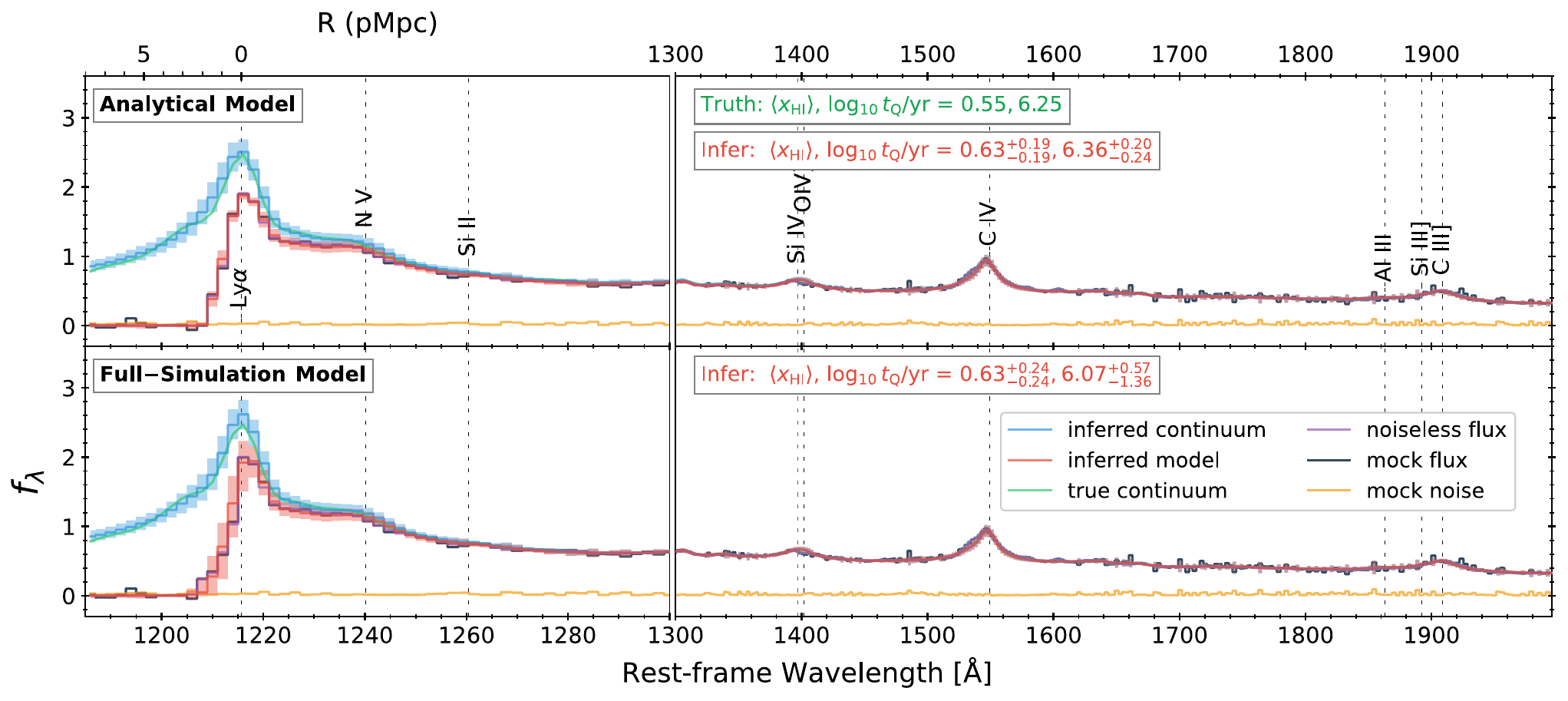}
    \caption{Model components of the true and the inferred spectrum of one of our $1000$ mock quasars considered in Section~\ref{sec:error_budget} with a global IGM neutral fraction of $\langle x_\mathrm{HI} \rangle = 0.55$ and a quasar lifetime of $\log_{10} t_\mathrm{Q}/\mathrm{yr} = 6.25$. In the upper panel, IGM transmission skewers are generated by the analytical model introduced in Section~\ref{sec:analytic}, in the lower panel by the full-simulation model from Section~\ref{sec:sims}. The PCA continuum model has a latent dimension of $n_\mathrm{latent} = 5$ and the observational setup is fixed to $\lambda_\mathrm{cut} = 2000\,\text{\AA}$, $\mathrm{S}/\mathrm{N} = 10$, and $\mathrm{FWHM} = 100\;\mathrm{km}/\mathrm{s}$ rebinned to $\mathrm{d}v_\mathrm{blue} = 500\;\mathrm{km}/\mathrm{s}$. For better visibility of the damping wing region, the rest-frame wavelength axis is stretched on the left side of the figure between $1185\,\text{\AA}$ and $1300\,\text{\AA}$. The full mock spectrum of the quasar is depicted in black and consists of the true continuum (green) combined with IGM transmission (purple) and mock spectral noise (yellow). The inferred model spectrum is depicted in red, consisting of the inferred continuum (blue) combined with the inferred IGM transmission field. Solid lines represent the median inferred models, shaded regions denote the $16\,\%$ and the $84\,\%$ percentile variations reflecting parameter uncertainty, continuum reconstruction errors, as well as spectral noise. All inferred model spectra are simulated based on reweighted HMC samples, assuring that each ensemble passes a marginal coverage test with respect to the two astrophysical parameters $\langle x_\mathrm{HI} \rangle$ and $t_\mathrm{Q}$. The $1\sigma$-uncertainties are notably larger for the full-simulation model due to the additional stochasticity that it encompasses.}
    \label{fig:analytic_vs_sims_spec}
\end{figure*}

\begin{figure}
    \centering
	\includegraphics[width=\columnwidth]{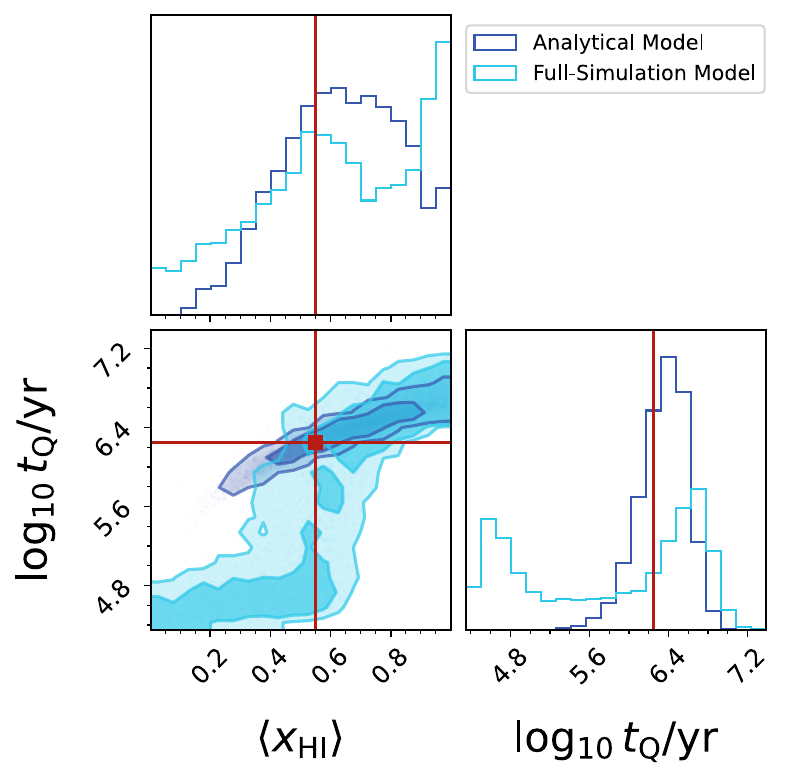}
    \caption{Marginal $(\langle x_\mathrm{HI} \rangle, t_\mathrm{Q})$-posterior distributions inferred from the two mock spectra depicted in Figure~\ref{fig:analytic_vs_sims_spec}. All HMC samples are reweighted, assuring that each ensemble passes a marginal coverage test with respect to the two astrophysical parameters $\langle x_\mathrm{HI} \rangle$ and $t_\mathrm{Q}$ as described in Section~\ref{sec:inf_tests}. The contours of the full-simulation model (green) extend notably farther along the axis of degeneracy between global IGM neutral fraction and quasar lifetime than those of the deterministic analytical model (blue).}
    \label{fig:analytic_vs_sims_contour}
\end{figure}

While our inference scheme can be readily applied to real observational data, we restrict this work to mock spectra in order to quantify the error budget of the inference pipeline. As in \citet{hennawi2024}, our models resemble the quasar ULAS J1342+0928 at $z=7.54$ with $J=20.3$ and realistic forward modelling of (hypothetical) instrument systematics and heteroscedastic spectral noise.

We generate mock spectra $\boldsymbol{f}$ by combining unabsorbed continua $\boldsymbol{s}$ with IGM transmission fields $\boldsymbol{t}$ and a noise realization $\boldsymbol{\sigma}$. To that end, we draw real low-redshift continua $\boldsymbol{s}$ from the set of $778$ autofit test objects introduced in Section~\ref{sec:dim_red},
and random spectral amplitudes $s_\mathrm{norm}$ from a uniform distribution, normalizing the continuum flux at $1450\,\text{\AA}$ to the drawn value of $s_\mathrm{norm}$. The range of the spectral amplitude distribution is chosen in accordance with the minimum and maximum amplitudes among $100$ calibration spectra (with magnitudes in the range $J = 20.3 \pm 0.1$).

To obtain IGM transmission fields $\boldsymbol{t}$, we draw realizations of the astrophysical parameters $\boldsymbol{\theta} = (\langle x_\mathrm{HI} \rangle, t_\mathrm{Q})$, and then apply one of the forward models introduced in Sections~\ref{sec:sims} and \ref{sec:analytic} to obtain a realization of $\boldsymbol{t}$ at the drawn parameter values. The parameter ranges covered by these models also define implicit priors for the inference.
We adopt uniform distributions for both astrophysical parameters (uniform in log-space for $t_\mathrm{Q}$) and choose the limits such that they cover a broad range of values that might be observationally relevant, noting that previous analysis of lower-redshift data has hinted towards a significantly more narrow log-normal quasar lifetime distribution centered around $t_\mathrm{Q} \sim 10^6\,\mathrm{yr}$ \citep{Khrykin2021}.

We consider global IGM neutral fractions in the full range $\langle x_\mathrm{HI} \rangle \in [0, 1]$. When working with the analytical IGM transmission model, we have to restrict this to $\langle x_\mathrm{HI} \rangle \in [10^{-4}, 1]$ to avoid divergences in our model (see Eq.~(\ref{eq:R_ion})). The quasar lifetime range in the full-simulation model is $t_\mathrm{Q}\in[10^3\,\mathrm{yr}, 10^8\,\mathrm{yr}]$ in accordance with our simulation grid. For the analytical model, we allow for lifetimes between $t_\mathrm{Q} \in [10^5\;\mathrm{yr}, 10^{8.84}\;\mathrm{yr}]$, where the upper lifetime limit corresponds to the age of the universe at redshift $z_\mathrm{QSO}=7.54$. Whenever we directly compare the two models, we adjust the lifetime range of the analytical model to that of the full-simulation model.

Note also that each of the $1183$ skewers governing the IGM transmission field $\boldsymbol{t}$ in our full-simulation model is pre-computed on a discrete $21\times 51$ grid of parameter values $\boldsymbol{\theta} = (\langle x_\mathrm{HI} \rangle, t_\mathrm{Q})$. We randomly choose one of the $1183$ available skewers and linearly interpolate the IGM transmission field $\boldsymbol{t}$ along the parameter grid to the drawn value of $\boldsymbol{\theta}$.

To simulate instrumental effects, we subsequently convolve the IGM transmission field $\boldsymbol{t}$e with a Gaussian line-spread function (LSF) of a given width (by default $\mathrm{FWHM} = 100 \,\mathrm{km}/\mathrm{s}$) and rebin it onto the final velocity grid. 
Since the transmission field is defined only on a rest-frame wavelength grid of $1185-2000\,\text{\AA}$, whereas our PCA model operates on the range $1170-2040\,\text{\AA}$, we have to interpolate the continuum onto the smaller grid in order to combine it with the transmission skewers. To keep the computational cost of the inference procedure at a tractable minimum, we work on a hybrid wavelength grid by splitting up the spectrum at $\lambda_\mathrm{blue-red} = 1218.10\,\text{\AA}$ into a high-resolution blue part and a low-resolution red part as described in \citet{hennawi2024}. 
High resolution is only relevant around the Lyman-$\alpha$ line and blueward in the Lyman-$\alpha$ forest region, while a coarse wavelength grid is sufficient to resolve the smooth part of the spectrum redward of $\lambda_\mathrm{blue-red}$. Throughout the analysis, we restrict the red part to a relatively coarse velocity pixel spacing ($\mathrm{d} v_\mathrm{red} = 500\,\mathrm{km}/\mathrm{s}$) and only investigate the effects of resolving the Lyman-$\alpha$ region with varying resolution. 
When we change spectral resolution, we always keep the spectral sampling factor between the $\mathrm{FWHM}$ of the LSF and the velocity spacing $\mathrm{d} v_\mathrm{blue}$ in the blue part of the spectrum fixed at $\mathrm{FWHM}/\mathrm{d} v_\mathrm{blue}=2$.

Lastly, we use \texttt{SkyCalc\_ipy} \citep{SkyCalc_ipy2021}
to generate realistic noise realizations including telluric absorption and contributions from object photons, sky background, and detector read noise. We tune the exposure time of our hypothetical instrument such that we achieve a given median signal-to-noise ratio $\mathrm{S}/\mathrm{N}$ per $100\,\mathrm{km}/\mathrm{s}$ velocity interval (by default $\mathrm{S}/\mathrm{N} = 10$).  %

For the purposes of our sensitivity analysis, we work with a fixed sample of $100$ mock spectra drawn from the priors introduced above. Our pipeline is implemented in such a way that we can generate the exact same set of mock objects 
given different (hyper-)parameter settings such as signal-to-noise ratio $\mathrm{S}/\mathrm{N}$ or spectral resolution $\mathrm{FWHM}$. Note that this includes the draw of the continuum as well as the Lyman-$\alpha$ transmission skewers, and, in all cases besides spectral resolution, also the noise realization. This allows us to isolate the impact of the aforementioned hyper-parameters on inference precision from the stochasticity of the process itself by considering a moderate number of $100$ mock spectra.

\subsection{Inference Procedure}
\label{sec:inference}

The likelihood derived in Section~\ref{sec:methods} in combination with the forward models from Sections~\ref{sec:sims} and \ref{sec:analytic} allows us to perform Bayesian inference to compute the posterior distribution of parameters $\boldsymbol{\Theta} = (\langle x_\mathrm{HI} \rangle, \log_{10} t_\mathrm{Q}/\mathrm{yr}, s_\mathrm{norm}, \boldsymbol{\eta})$ given a (mock) observational quasar spectrum $\boldsymbol{f}$. Again, $(\langle x_\mathrm{HI} \rangle, \log_{10} t_\mathrm{Q}/\mathrm{yr})$ are the two astrophysical parameters of interest that we infer jointly with the continuum nuisance parameters $\boldsymbol{\eta} = (s_\mathrm{norm}, \boldsymbol{\xi})$, i.e., the continuum amplitude and the best-fitting PCA coefficients. For the purposes of this paper, we adopt uniform priors on the astrophysical parameters $\langle x_\mathrm{HI} \rangle$ and $\log_{10} t_\mathrm{Q}/\mathrm{yr}$ that arise from the fixed parameter grid in the full-simulation model or the analytical model domain. The respective parameter ranges are summarized in Section~\ref{sec:mock}. Our prior on the spectral amplitude $s_\mathrm{norm}$ is $\mathrm{Uniform}(0, \infty)$, assuring positivity of this parameter, and we allow each PCA coefficient $\alpha_i$ of the continuum model to vary over the full range $\mathrm{Uniform}(-\infty, \infty)$. Practically, in our numerical pipeline, we apply variable transformations to all (half-)bounded parameters to make them fully unbounded.

We sample from the posterior distribution using a Hamiltonian Monte-Carlo (HMC) algorithm with a No U-Turn Sampler (NUTS) kernel implemented in the \texttt{NumPyro} probabilistic programming package. The \texttt{NumPyro} package is based on the machine learning and autograd framework \texttt{JAX}, allowing us to implement our scheme in a fully differentiable and GPU-accelerated way. For each inference task, we run $4$ MCMC chains for $2000$ MCMC steps each, $1000$ of which are discarded as warm-up steps.

An example
inference run is shown in Figure~\ref{fig:analytic_vs_sims_spec}. The mock object is part of the ensemble described and analyzed in more detail in Section~\ref{sec:error_budget}. We compare in the upper and the lower panel the same object, combined with full-simulation and analytical IGM transmission skewers as introduced in Sections~\ref{sec:sims} and \ref{sec:analytic}, respectively. Note that the shaded $1\sigma$ regions need to entail the uncertainty contributions from all individual model components to allow for a meaningful assessment of the goodness-of-fit of the inferred model. To that end, we compute for each HMC sample $(\boldsymbol{\theta}, \boldsymbol{\eta})$ a draw of the continuum $\boldsymbol{s} = \boldsymbol{s}_\mathrm{DR}(\boldsymbol{\eta})\circ(\boldsymbol{1}+\boldsymbol{\delta})$, where we obtain $\boldsymbol{s}_\mathrm{DR}(\boldsymbol{\eta})$ by evaluating our PCA model and the spectral amplitude at $\boldsymbol{\eta}$, and obtain $\boldsymbol{\delta}$ by drawing a realization of the associated continuum reconstruction error from Eq.~(\ref{eq:P_s_alpha}). We further draw a random IGM transmission skewer $\boldsymbol{t}$ for the given value of $\boldsymbol{\theta}$, using the analytical or the full-simulation IGM transmission model, respectively, giving $\boldsymbol{f} = \boldsymbol{s}\circ\boldsymbol{t}$. Lastly, we draw a noise realization $\tilde{\boldsymbol{\sigma}}$ from a Gaussian distribution with noise vector $\boldsymbol{\sigma}$. The depicted uncertainties on the inferred model and continuum are the $16\,\%$ and $84\,\%$ percentile variations of the realizations $\boldsymbol{f}+\tilde{\boldsymbol{\sigma}}$ and $\boldsymbol{s}+\tilde{\boldsymbol{\sigma}}$, hence reflecting parameter uncertainty, continuum reconstruction errors, as well as spectral noise. Note also that we use the coverage-corrected HMC samples for this procedure as discussed in Section~\ref{sec:inf_tests}.

While we are able to reconstruct the continuum reasonably well with either IGM transmission model, the error bars of the inferred model (red shaded regions) are significantly smaller when working with the analytical one. This is a direct consequence of the additional stochasticity present 
in the full-simulation model. As a result, also the marginal posterior distribution in Figure~\ref{fig:analytic_vs_sims_contour} extends significantly farther along the axis of degeneracy between $\langle x_\mathrm{HI} \rangle$ and $\log_{10} t_\mathrm{Q}/\mathrm{yr}$.

This degeneracy arises because both parameters affect the neutral hydrogen content in the surrounding IGM and thus have a direct impact on the strength of the IGM damping wing signature. As a result, we can not always tell with certainty if we are observing a young quasar residing in a predominantly ionized IGM, or an older object in a globally more neutral IGM. This is because by carving out a larger ionized region, older quasars have significantly decreased the HI content along their sightlines as compared to the global volume-averaged value $\langle x_\mathrm{HI} \rangle$, making their transmission profiles appear more akin to that seen for younger objects which have carved out a shorter ionization front but reside in a globally more ionized IGM.

Since Eqs.~(\ref{eq:R_ion}), (\ref{eq:tau_DW}) and (\ref{eq:tau_GP}) directly encode these dependencies, Figure~\ref{fig:analytic_vs_sims_contour} shows a tight axis of degeneracy for the posterior obtained with the analytical IGM transmission model. In case of the full-simulation model, this degeneracy extends significantly farther across parameter space and entails notably more scatter due to the additional sources of stochasticity that are treated in this model: firstly, because of the patchiness of reionization, the HI content along a given sightline can differ significantly from the global volume-averaged value $\langle x_\mathrm{HI} \rangle$ which is approximated as uniform in the analytical IGM transmission model. Secondly, fluctuations in the density field are giving rise to additional stochasticity in the location of the ionization front in our full-simulation model. Altogether, this amounts to the notable degeneracy seen in Figure~\ref{fig:analytic_vs_sims_contour}.

In Section~\ref{sec:error_budget} we provide a quantitative analysis of how the differences between the two models impact the overall inference precision we can achieve.

\section{Sensitivity Analysis}
\label{sec:sensitivity}

Before applying our inference scheme to real-world data, it is essential to understand its sensitivity with respect to both model and hyper-parameters, as well as the associated error budget. The ultimate goal of the inference procedure is to obtain a 
1) \textit{faithful}, and 2) \textit{precise} estimate of the posterior distribution of the astrophysical parameters of interest, i.e., the global volume-averaged IGM neutral fraction $\langle x_\mathrm{HI} \rangle$ and the lifetime $t_\mathrm{Q}$ of the quasar at hand. In the next section, we will review the concept of coverage probability as a metric of overconfidence in the inferred posterior distribution and describe how to correct for potential imperfections
if required. Subsequently, we can introduce a \textit{precision} metric that we can use to understand the sensitivity of the inference scheme with respect to various hyper-parameters related to the continuum DR model and the observational setup of the input spectra.

In order to isolate their impact from additional sources of stochasticity, we consider the same set of $100$ mock spectra generated with different hyper-parameter choices as described in Section~\ref{sec:mock} throughout the entire analysis. Further, we eliminate the stochastic dependence of proximity zone size and IGM damping wing strength on the astrophysical parameters $\langle x_\mathrm{HI} \rangle$ and $t_\mathrm{Q}$ by adopting the analytical IGM transmission model introduced in Section~\ref{sec:analytic} as our default forward model of IGM transmission. Its deterministic relations (\ref{eq:R_ion}) and (\ref{eq:tau_DW}) allow us to identify the impact of the aforementioned hyper-parameters on inference precision as clearly as possible. In Appendix~\ref{app:param_scan_sims}, we show that the same trends, though weakened by the additional source of stochasticity, remain true when using realistic IGM transmission skewers based on cosmological simulations.

In the final part of this section, we will then, informed about the optimal hyper-parameter setting, provide a rigorous break-down of the contributions from all individual model components to the total error budget as a function of astrophysical parameter space.

\subsection{Inference Tests}
\label{sec:inf_tests}

We show that the inferred posterior distributions are statistically faithful 
by running inference tests for ensembles of mock objects drawn from the full prior range. By computing expected coverage probabilities, we confirm that the inferred posterior distribution
faithfully represents the true distribution of mock objects, i.e., that the inference is not under- or overconfident with regards to the inferred parameter values, or, in other words, that the inferred posterior distributions are not too wide and not too narrow. To this end, we introduce the concept of coverage probabilities: given 
a credibility level $\alpha \in [0, 1]$, the coverage probability $C_\alpha$ quantifies how frequently the true probability is contained in the $\alpha$-th credibility level of the inferred posterior \citep[see e.g.][]{Sellentin2019}. 

More practically speaking, for a given inference run we compute the credibility level $\alpha$ as the $\alpha$-th highest-density region (HDR) of the inferred posterior distribution by sorting all MCMC samples with respect to posterior probability and choosing the $\alpha$-th highest fraction. We then test if the true posterior value is contained in this $\alpha$-th HDR and repeat the procedure for each object in our ensemble of mock spectra. The fraction of objects for which this is the case is called the expected coverage probability $C_\alpha$.
In the ideal case, we should find $C_\alpha = \alpha$ at all coverage levels. In this case, we say that we \textit{pass} the coverage test. Otherwise, if $C_\alpha > \alpha$, we are said to be \textit{underconfident} or \textit{conservative} about the significance of our results, i.e., the inferred posterior distribution is wider than it could be according to the true distribution. On the other hand, and more crucially, if $C_\alpha < \alpha$, we are \textit{overconfident} about the significance of our results and our posterior distribution is shifted or narrower than the true distribution would allow.

In the latter case, it is necessary to broaden the posteriors to make them compatible with the true distribution. \citet{hennawi2024} \citep[see also][]{Wolfson2023} described a principled mathematical way of assigning unique weights to the draws of a MCMC chain to weigh them in such a way that the inferred coverage probability $C_\alpha$ exactly fulfills $C_\alpha = \alpha$ at all credibility levels $\alpha$. We apply these corrections to all runs considered in this paper as many runs show some degree of overconfidence (see Appendix~\ref{app:inf_tests}). For consistency, we also apply the same technique to runs that are consistent with perfect coverage as this does not affect the results and obviates the need to introduce a precise threshold for overconfidence.

Inference tests can be performed by considering the full-dimensional posterior distribution, or, alternatively, a lower-dimensional marginalized version of it. In our case, the only physical parameters of interest are the global IGM neutral fraction $\langle x_\mathrm{HI} \rangle$ and the quasar lifetime $t_\mathrm{Q}$, whereas the remaining $n_\mathrm{latent}+1$ parameters are nuisance parameters related to the shape of the continuum. In order to make sure that our posteriors are compatible not only with the distribution in full parameter space but also in the astrophysical subspace, we perform coverage tests with respect to both the full $(2+n_\mathrm{latent}+1)$-dimensional posterior distribution and the $2$-dimensional marginal $(\langle x_\mathrm{HI} \rangle, \log_{10} t_\mathrm{Q}/\mathrm{yr}\})$-posterior density. It turns out that the full coverage probability is always on par with or more conservative than the marginal one. We therefore adopt the marginal coverage probability as the decisive measure of overconfidence and use it to compute the above mentioned importance weights. We confirmed that in all cases considered that the marginally coverage corrected models also pass full coverage tests. We show the (marginal) inference test results for all mock ensembles treated in Section~\ref{sec:param_scan} in Appendix~\ref{app:inf_tests}.

\subsection{Inference Precision}
\label{sec:inf_prec}

After assuring that the inferred posteriors are unbiased in the sense defined above, our metric for inference precision is the width of the posterior distributions after marginalizing over all nuisance parameters $\boldsymbol{\eta}$ related to the quasar continuum, i.e., the PCA coefficients $\boldsymbol{\xi}$ and the amplitude of the spectrum $s_\mathrm{norm}$.
We adopt as a measure for this width the (symmetized) $1\sigma$-error $\Delta_{68\,\%}(x)$ of the parameter $x\in\{\langle x_\mathrm{HI} \rangle, \log_{10} t_\mathrm{Q}/\mathrm{yr}\}$, i.e., half the distance between the $84\,\%$ and $16\,\%$ percentile of the respective $1\mathrm{d}$-marginal posterior density
\begin{equation}
\label{eq:prec}
    \Delta_{68\,\%}(x) = \frac{1}{2} (P_{84\,\%}(x) - P_{16\,\%}(x)),
\end{equation}
where $P_{q}(x)$ is the $q$-th percentile of the marginal posterior with respect to the parameter $x$.

We quantify precision in a statistical sense by averaging over the inference results of the mock ensemble with $\langle x_\mathrm{HI} \rangle$ and $t_\mathrm{Q}$ drawn from the full prior volume as defined in Section~\ref{sec:mock}. Given that our method will mainly target the transition from a neutral to a reionized universe at intermediate neutral fractions, and that current measurements favor a significantly more narrow quasar lifetime distribution than that covered by our prior \citep{Khrykin2021}, it is instructive to investigate how well our inference scheme performs in this particular region of parameter space. To that end, we define a subset of \textit{fiducial} objects with neutral fractions in the range $0.25 \leq \langle x_\mathrm{HI} \rangle \leq 0.75$ and lifetimes between $5.25 \leq \log_{10} t_\mathrm{Q}/\mathrm{yr} \leq 6.75$. In addition to the overall median precision, we always quote the median of these 16 (11) objects for the analytical (full-simulation) IGM transmission model to confirm that the trends we identify are not the result 
of an atypical 
region of parameter space.

\subsection{Parameter Scan}
\label{sec:param_scan}

\begin{figure*}
    \centering
	\includegraphics[width=\textwidth]{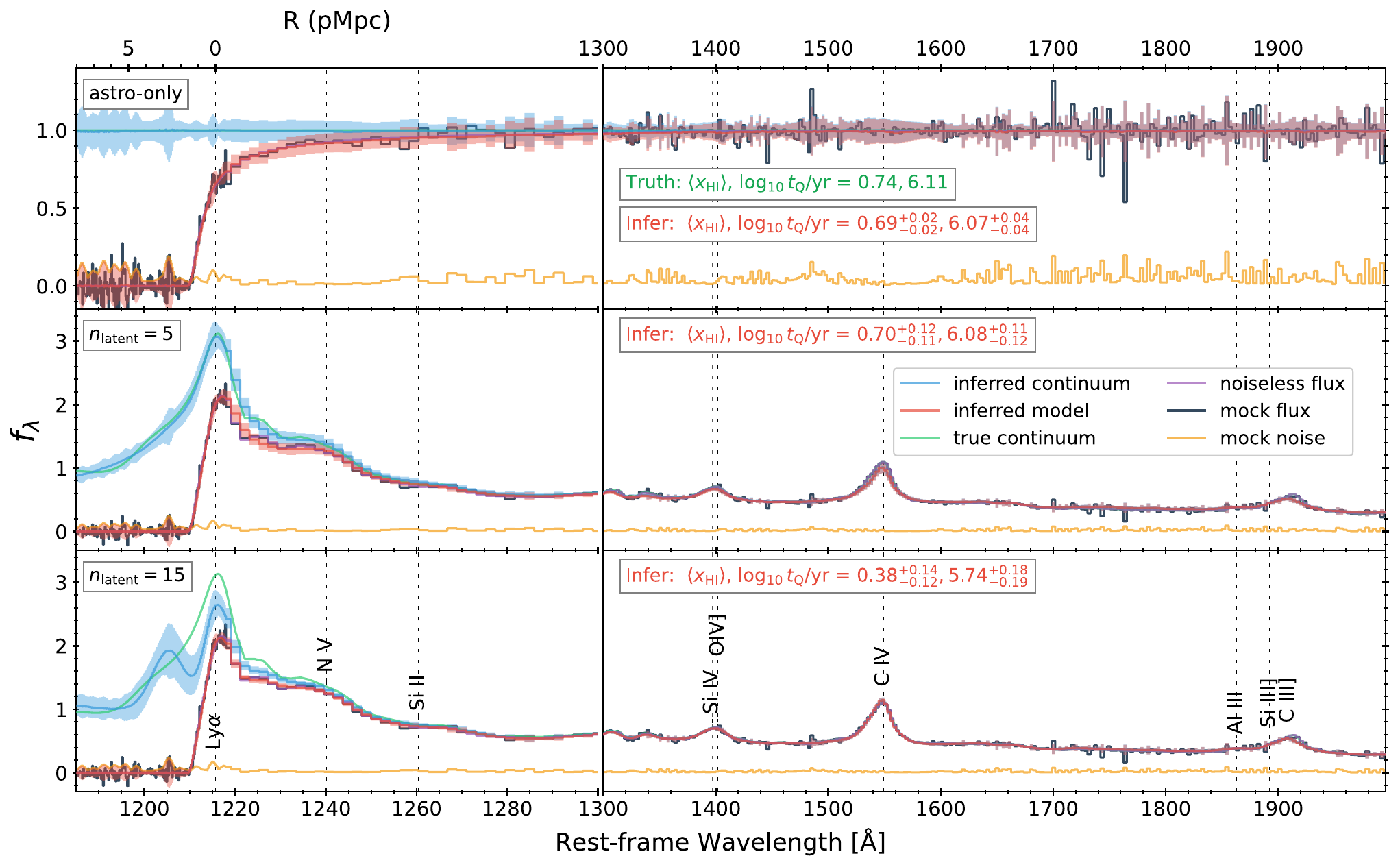}
    \caption{Model components of the true and the inferred spectrum of one of our $100$ mock quasars with $\langle x_\mathrm{HI} \rangle = 0.74$ and $\log_{10} t_\mathrm{Q}/\mathrm{yr} = 6.11$, considered with different latent dimensions $n_\mathrm{latent}$ for the PCA continuum model. IGM transmission skewers are generated by the analytical model introduced in Section~\ref{sec:analytic}, and the observational setup is fixed to $\lambda_\mathrm{cut} = 2000\,\text{\AA}$, $\mathrm{S}/\mathrm{N} = 10$ and $\mathrm{FWHM} = 100\;\mathrm{km}/\mathrm{s}$. For better visibility of the damping wing region, the rest-frame wavelength axis is stretched on the left side of the figure between $1185\,\text{\AA}$ and $1300\,\text{\AA}$. The full mock spectrum of the quasar is depicted in black and consists of the true continuum (green) combined with IGM transmission (purple) and mock spectral noise (yellow). The inferred model spectrum is depicted in red, consisting of the inferred continuum (blue) combined with the inferred IGM transmission field. Solid lines represent the median inferred models, shaded regions denote the $16\,\%$ and the $84\,\%$ percentile variations reflecting parameter uncertainty, continuum reconstruction errors, as well as spectral noise. All inferred model spectra are simulated based on reweighted HMC samples, assuring that each ensemble passes a marginal coverage test with respect to the two astrophysical parameters $\langle x_\mathrm{HI} \rangle$ and $t_\mathrm{Q}$. The top panel shows the results for a continuum-normalized version of the spectrum, i.e., theoretically optimal results assuming perfect knowledge of the continuum. The middle and the bottom panel show runs with $n_\mathrm{latent} = 5$ and $n_\mathrm{latent} = 15$, respectively. The inferred continuum of the $n_\mathrm{latent} = 15$ model is subject to significantly more uncertainty than that of the lower dimensional model, indicating that additional PCA vectors do not encode relevant physical information about the Lyman-$\alpha$ forest or the damping wing region.}
    \label{fig:lat_dim_spec}
\end{figure*}

\begin{figure*}
    \centering
	\includegraphics[height=0.45\textwidth]{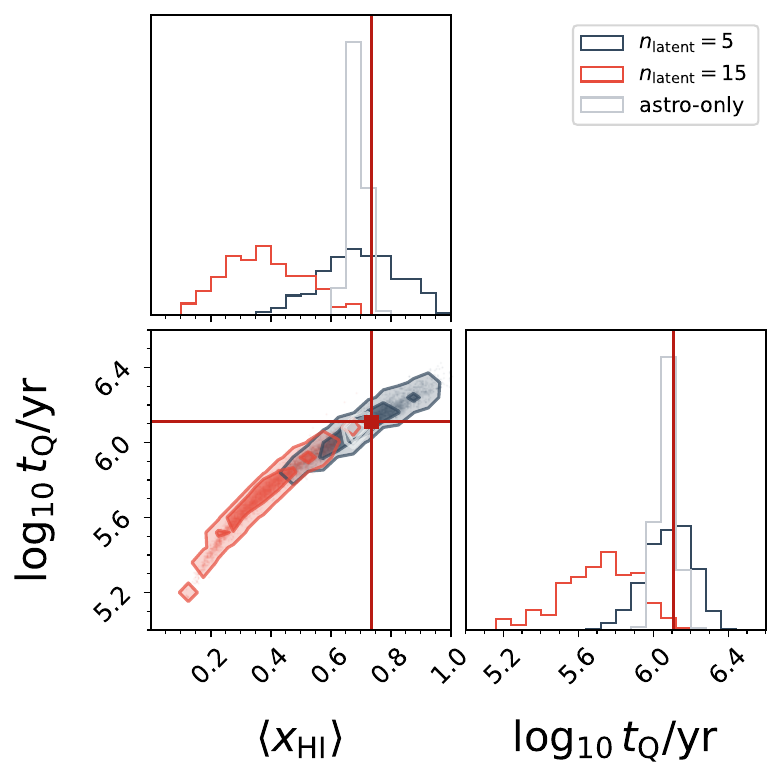} \hspace{1cm}
	\includegraphics[height=0.45\textwidth]{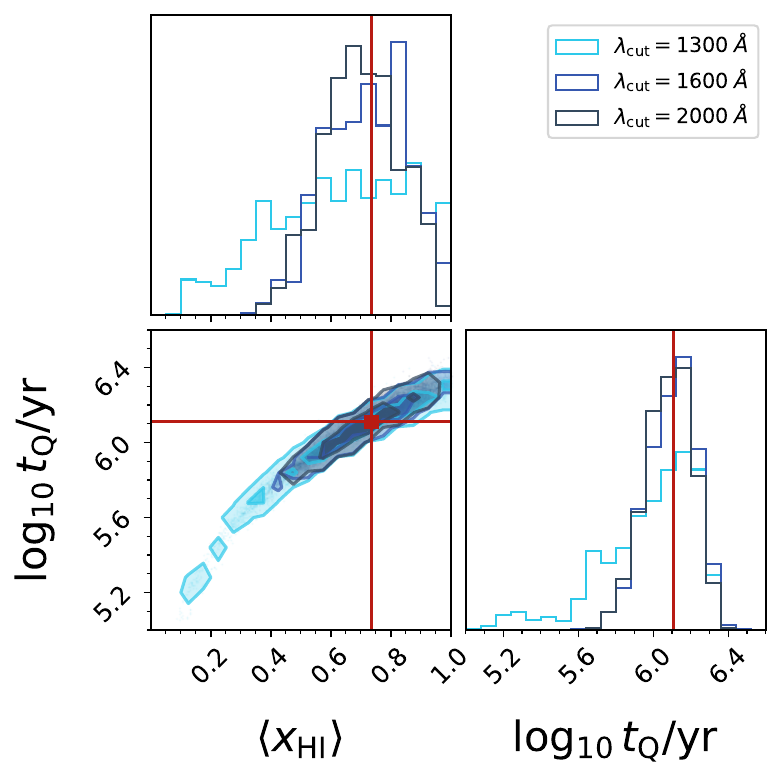} \vspace{0.5cm} \\
	\includegraphics[height=0.45\textwidth]{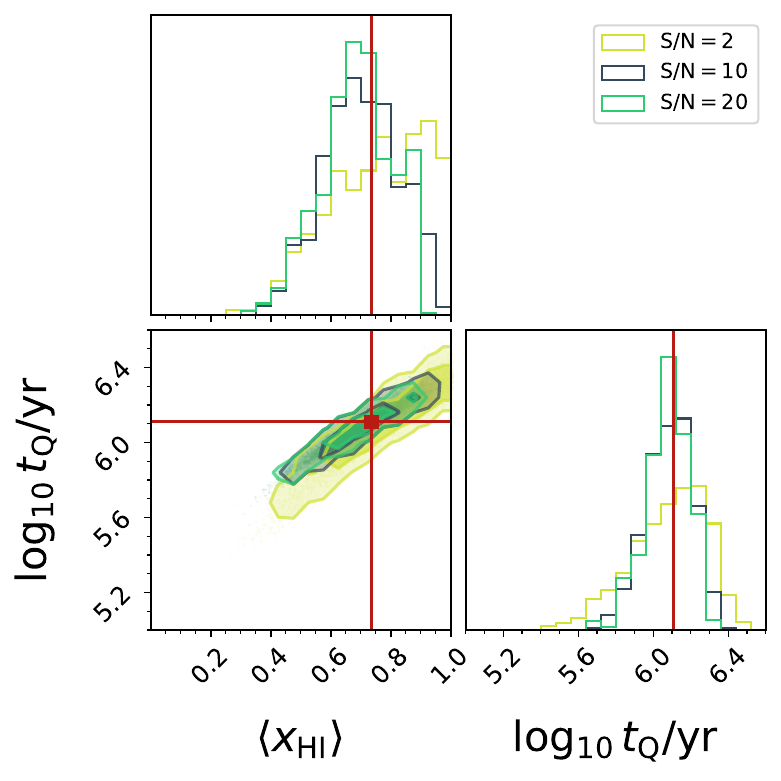} \hspace{0.9cm}
	\includegraphics[height=0.45\textwidth]{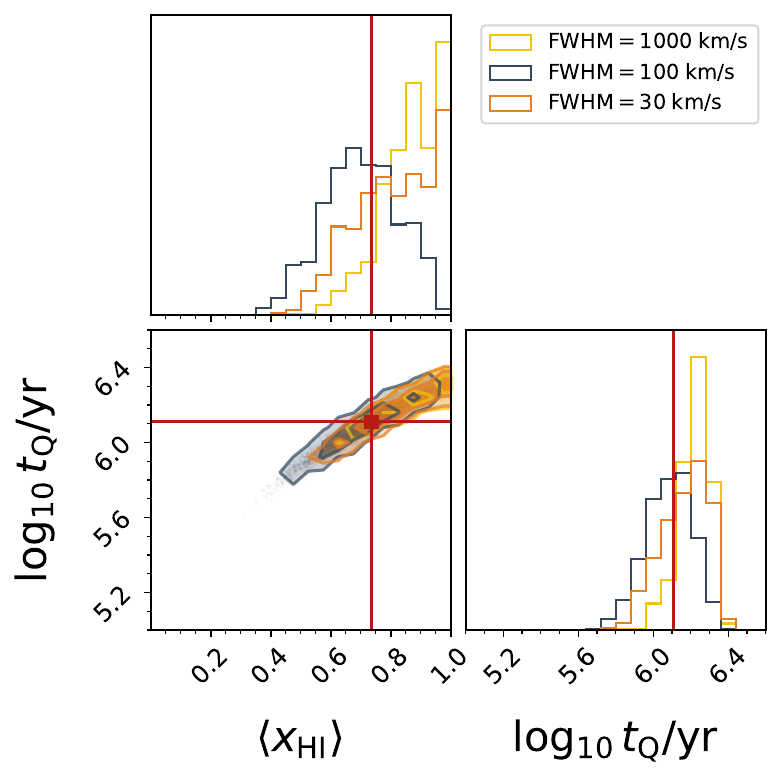}
    \caption{Marginal $(\langle x_\mathrm{HI} \rangle, t_\mathrm{Q})$-posterior distributions inferred from one of our $100$ mock quasars with $\langle x_\mathrm{HI} \rangle = 0.74$ and $\log_{10} t_\mathrm{Q}/\mathrm{yr} = 6.11$ with different hyper-parameter settings. IGM transmission skewers are generated by the analytical model introduced in Section~\ref{sec:analytic}. All HMC samples are reweighted, assuring that each ensemble passes a marginal coverage test with respect to the two astrophysical parameters $\langle x_\mathrm{HI} \rangle$ and $t_\mathrm{Q}$. Whenever applicable, we marginalized over all nuisance parameters related to the shape of the continuum. We vary latent dimension $n_\mathrm{latent}$, red-side wavelength coverage $\lambda_\mathrm{cut}$, signal-to-noise ratio $\mathrm{S}/\mathrm{N}$, and spectral resolution $\mathrm{FWHM}$, in the four panels from left to right. Each panel contains three representative choices of the respective parameter. The corresponding inferred models are shown in Figures~\ref{fig:lat_dim_spec} and \ref{fig:mix_spec}.
    Our reference run with $n_\mathrm{latent} = 5$, $\lambda_\mathrm{cut} = 2000\,\text{\AA}$, $\mathrm{S}/\mathrm{N} = 10$ and $\mathrm{FWHM} = 100\;\mathrm{km}/\mathrm{s}$) is depicted in black in each panel. The latent dimension plot also shows the theoretically optimal posteriors as obtained from the run with continuum-normalized spectra labelled as 'astro-only'.}
    \label{fig:mix_contour}
\end{figure*}

\begin{figure*}
    \centering
	\includegraphics[width=0.98\textwidth]{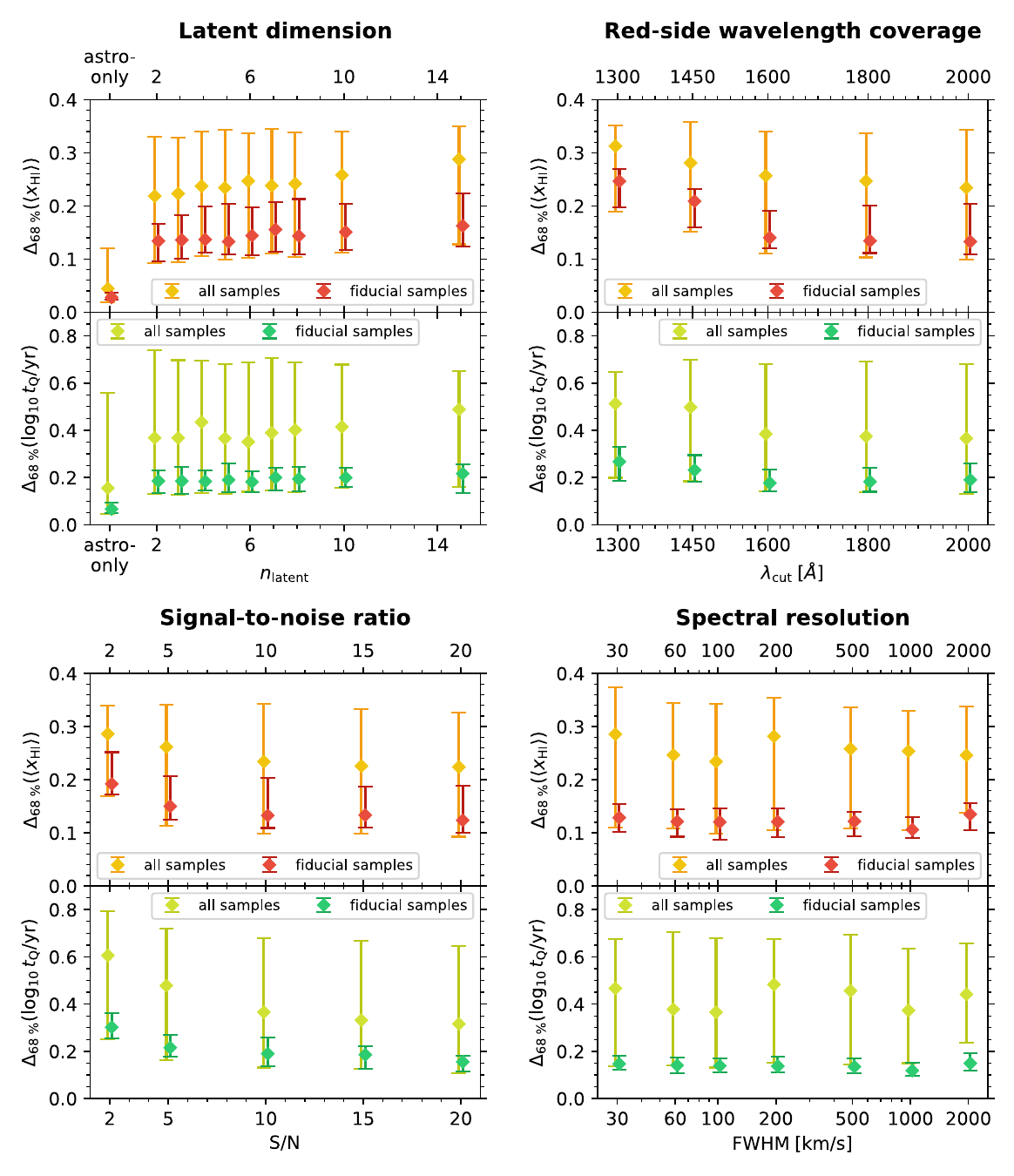}
    \caption{Inference precision with respect to IGM neutral fraction $\langle x_\mathrm{HI} \rangle$ (upper half-panels) and quasar lifetime $t_\mathrm{Q}$ (lower half-panels) as a function of different hyper-parameters (i.e., $n_\mathrm{latent}$, $\lambda_\mathrm{cut}$, $\mathrm{S}/\mathrm{N}$, and $\mathrm{FWHM}$). IGM transmission skewers are generated by the analytical model introduced in Section~\ref{sec:analytic}. %
    Median precision values with respect to each ensemble of $100$ mock quasars are depicted in yellow (light green), and with respect to the subset of $16$ fiducial spectra in red (green), shifted slightly in $x$-direction with respect to each other for visibility. Error bars correspond to the $16\,\%$ and $84\,\%$ percentiles of the respective mock ensemble. In the case of spectral resolution, the fiducial samples are taken from a separate set of $100$ objects with fixed neutral fraction $\langle x_\mathrm{HI} \rangle = 0.5$ and fixed lifetime $t_\mathrm{Q} = 10^6\,\mathrm{yr}$ to reduce stochasticity arising from varying noise draws between mocks at different resolutions. Labelled as 'astro-only' on the left of the latent dimension plot, we show the optimal precision bounds as obtained from the run with continuum-normalized spectra. All precision values are based on reweighted marginal posterior PDFs, assuring that each mock ensemble passes a marginal coverage test with respect to the two astrophysical parameters $\langle x_\mathrm{HI} \rangle$ and $t_\mathrm{Q}$.} %
    \label{fig:mix_precision}
\end{figure*}

\begin{figure*}
    \centering
	\includegraphics[width=\textwidth]{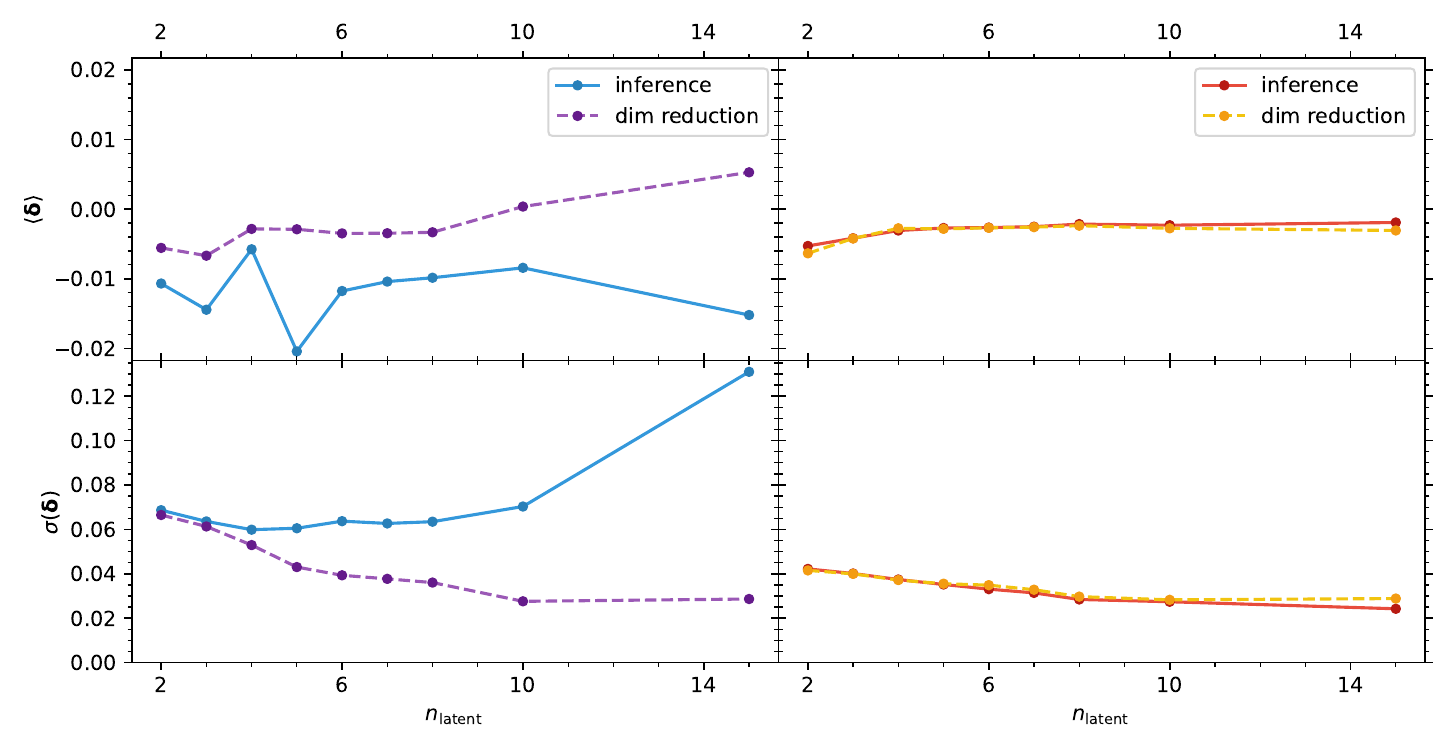}
    \caption{Moments of the continuum reconstruction error $\boldsymbol{\delta}$ as a function of latent dimension $n_\mathrm{latent}$ of the PCA continuum model. The panels on the left show the mean $\langle \boldsymbol{\delta} \rangle$ and the standard deviation $\sigma(\boldsymbol{\delta})$ of the continuum reconstruction error with respect to the blue pixel range ($1185\,\text{\AA} \leq \lambda \leq 1260\,\text{\AA}$). The panels on the right depict the same moments with respect to the red pixels of the spectrum ($1260\,\text{\AA} < \lambda \leq 2000\,\text{\AA}$). Each individual data point represents the median of the corresponding moment with respect to the $100$ object mock ensemble with $\lambda_\mathrm{cut} = 2000\,\text{\AA}$, $\mathrm{S}/\mathrm{N} = 10$ and $\mathrm{FWHM} = 100\;\mathrm{km}/\mathrm{s}$. The moments related to the full inference task $\boldsymbol{\delta}_\mathrm{inf}$ are depicted as solid blue (red) lines, the ones arising solely from the dimensionality reduction task $\boldsymbol{\delta}_\mathrm{DR}$ as dashed purple (yellow) lines. The standard deviation $\sigma(\boldsymbol{\delta}_\mathrm{inf})$ of the blue-side reconstruction error of the inferred continuum has a shallow minimum around $n_\mathrm{latent} \simeq 4$, whereas $\sigma(\boldsymbol{\delta}_\mathrm{DR})$ keeps decreasing monotonically with latent dimension like in the smooth red part of the spectrum.}
    \label{fig:lat_dim_reconstr_err}
\end{figure*}

\begin{figure*}
    \centering
	\includegraphics[width=0.98\textwidth]{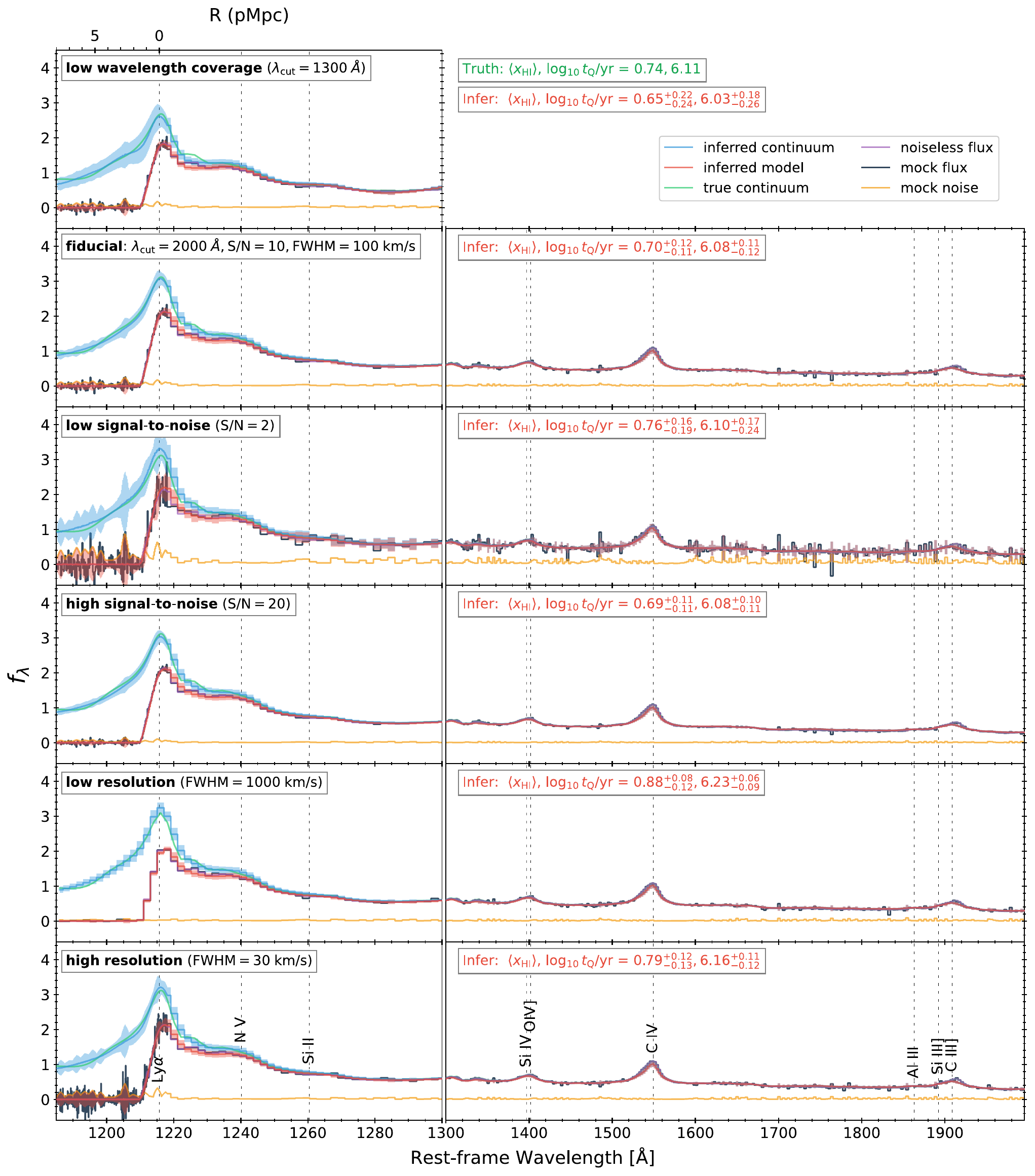}
    \caption{Model components of the true and the inferred spectrum of one of our $100$ mock quasars with $\langle x_\mathrm{HI} \rangle = 0.74$ and $\log_{10} t_\mathrm{Q}/\mathrm{yr} = 6.11$, considered in different observational setups, and compared to our reference setting of $\lambda_\mathrm{cut} = 2000\,\text{\AA}$, $\mathrm{S}/\mathrm{N} = 10$ and $\mathrm{FWHM} = 100\;\mathrm{km}/\mathrm{s}$. IGM transmission skewers are generated by the analytical model introduced in Section~\ref{sec:analytic}. For better visibility of the damping wing region, the rest-frame wavelength axis is stretched on the left side of the figure between $1185\,\text{\AA}$ and $1300\,\text{\AA}$. The full mock spectrum of the quasar is depicted in black and consists of the true continuum (green) combined with IGM transmission (purple) and mock spectral noise (yellow). The inferred model spectrum is depicted in red, consisting of the inferred continuum (blue) combined with the inferred IGM transmission field. Solid lines represent the median inferred models, shaded regions denote the $16\,\%$ and the $84\,\%$ percentile variations reflecting parameter uncertainty, continuum reconstruction errors, as well as spectral noise. All inferred model spectra are simulated based on reweighted HMC samples, assuring that each ensemble passes a marginal coverage test with respect to the two astrophysical parameters $\langle x_\mathrm{HI} \rangle$ and $t_\mathrm{Q}$. Lower red-side wavelength coverage or a lower signal-to-noise ratio increases the uncertainties of the inferred model, whereas the dependence on resolution is relatively weak.}
    \label{fig:mix_spec}
\end{figure*}

In this section, we will systematically investigate the effects of various hyper-parameters related to the continuum model and the observational setup 
on inference precision as measured in terms of $\Delta_{68\,\%}(\langle x_\mathrm{HI}\rangle)$ and $\Delta_{68\,\%}(\log_{10} t_\mathrm{Q}/\mathrm{yr})$ defined in Eq.~(\ref{eq:prec}). We do so by subsequently varying 1) the latent dimension $n_\mathrm{latent}$ of our PCA continuum model, 2) the maximum wavelength coverage $\lambda_\mathrm{cut}$ on the red side of the quasar spectrum, 3) the signal-to-noise ratio $\mathrm{S}/\mathrm{N}$ per $100\,\mathrm{km}/\mathrm{s}$ velocity stretch, and 
4) spectral resolution on the blue side of the spectrum, quantified in terms of $\mathrm{FWHM}$ of the LSF, keeping the spectral sampling factor fixed to $\mathrm{FWHM}/\mathrm{d}v_\mathrm{blue} = 2$. 
When varying one of these parameters, we keep the other parameters fixed to the fiducial values of $n_\mathrm{latent} = 5$, $\lambda_\mathrm{cut} = 2000\,\text{\AA}$, $\mathrm{S}/\mathrm{N} = 10$, and $\mathrm{FWHM} = 100\;\mathrm{km}/\mathrm{s}$ (with a velocity spacing of $\mathrm{d}v_\mathrm{blue} = 50\;\mathrm{km}/\mathrm{s}$).

\subsubsection{Latent Dimension}
\label{sec:lat_dim}

An important component of our pipeline is the quasar continuum dimensionality reduction (DR) model introduced in Section~\ref{sec:dim_red}. It reduces the number of pixels $n_\lambda = O(100-1000)$ covering the full spectral range to a low $n_\mathrm{latent}$-dimensional parametric (PCA) representation based on $14\,781$ low-redshift SDSS-autofit continua. The PCA vectors are rank-ordered by the amount of variation in the sample that they account for. By only keeping the first $n_\mathrm{latent}$ vectors (where $n_\mathrm{latent} \ll n_\lambda$), we can reduce the dimensionality of our continuum model to an arbitrary extent, typically $n_\mathrm{latent} = O(1-10)$. The precise number of vectors $n_\mathrm{latent}$ is a free hyper-parameter of the model. It should be large enough to allow for sufficient model flexibility, yet too high of a number of latent factors comes with the danger of overfitting. In this section we determine the optimal dimensionality of our PCA continuum model for the inference task at hand, which is to say the smallest latent dimension that minimizes the errors on the inferred astrophysical parameters $\langle x_\mathrm{HI}\rangle$ and $t_\mathrm{Q}$.

Figure~\ref{fig:lat_dim_spec} shows a representative example of our inference pipeline applied to the same quasar spectrum with different choices for the latent dimension of the continuum model. To analyze the overall contribution of the continuum model to the error budget, we also generated a continuum-normalized version of the spectrum. Normalizing the mock spectrum by the (in reality unknown) continuum completely removes the need for the continuum DR model. In this way, we eliminate all continuum nuisance parameters from the analysis (i.e., PCA coefficients $\boldsymbol{\xi}$ as well as the normalization factor $s_\mathrm{norm}$ of the continuum), leaving the astrophysical parameters $\langle x_\mathrm{HI} \rangle$ and $t_\mathrm{Q}$ as the only two parameters to infer. This setting can be seen as an optimal bound as to how constraining the resulting posteriors can theoretically get assuming perfect knowledge of the continuum. In Figure~\ref{fig:lat_dim_spec}, we compare such a run to runs with a low-dimensional continuum model with $n_\mathrm{latent} = 5$ and a higher-dimensional one with $n_\mathrm{latent} = 15$. 
The continuum inferred with $n_\mathrm{latent}  = 15$ shows an intense oscillation blueward of the Lyman-$\alpha$ line that is not present in the underlying mock continuum, suggesting that this model is too flexible and that the additional PCA vectors do not capture additional physics.

In the left panel of Figure~\ref{fig:mix_contour} we show the marginalized posterior distributions of $\langle x_\mathrm{HI} \rangle$ and $t_\mathrm{Q}$ for the three inference runs depicted in Figure~\ref{fig:lat_dim_spec}. Note that these and all subsequent runs have been coverage corrected according to the procedure described in Section~\ref{sec:inf_tests} to assure exact marginal coverage in the 2d-astrophysical parameter space. This assures that we do not misinterpret narrow but overconfident posteriors as overly precise. We show all original inference test results in Figure~\ref{fig:coverage_analytic} in Appendix~\ref{app:inf_tests}.

It is clearly apparent that the posterior PDF of the $n_\mathrm{latent} = 15$ model is significantly less informative and extends further along the axis of degeneracy between the two astrophysical parameters than the PDF of the lower-dimensional model. Independently of the choice of latent dimension, the posteriors are significantly wider than those obtained with the continuum-normalized model where both parameters are very tightly constrained, indicating that continuum reconstruction contributes significantly to the overall error budget.

These individual trends are confirmed when considering the median inference precision $\Delta_{68\,\%}(\langle x_\mathrm{HI} \rangle)$ and $\Delta_{68\,\%}(\log_{10} t_\mathrm{Q}/\mathrm{yr})$ as introduced in Section~\ref{sec:inf_prec} with respect to $100$ mock samples drawn from the full prior volume. We depict $\Delta_{68\,\%}(\langle x_\mathrm{HI} \rangle)$ and $\Delta_{68\,\%}(\log_{10} t_\mathrm{Q}/\mathrm{yr})$ with respect to both the full sample ($100$ objects) and the fiducial subset ($16$ objects) as a function of latent dimension in the upper left panels of Figure~\ref{fig:mix_precision}. The errorbars are estimated in terms of the $16\,\%$ and $84\,\%$-percentiles with respect to the respective set of objects. We observe that regardless of the choice of $n_\mathrm{latent}$, the posteriors are $2-3$ times wider than the theoretically optimal ones obtained in the continuum-normalized case. This indicates significant contributions of continuum reconstruction to the total error budget which we quantify in further detail at the end of this section.
Secondly, there is surprisingly little variation among different choices of latent dimension between $2 \lesssim n_\mathrm{latent} \lesssim 10$. If any, there is a slight trend of deteriorating inference precision with higher latent dimension, becoming more evident at $n_\mathrm{latent} \gtrsim 10$. We observe the same qualitative behavior for the fiducial subset of mock spectra.

In our companion paper \citep{hennawi2024}, we tested at which latent dimension $n_\mathrm{latent}$ the relative continuum reconstruction error flattens out when representing continua with a $n_\mathrm{latent}$-dimensional PCA continuum model. We found that the reconstruction error continued to decrease with dimension at least out to $n_\mathrm{latent} \sim 10$ \citep[cf. Figures~3 and 4 in][]{hennawi2024}. Note that here we are asking a different question, that is to say at which latent dimension the precision on the inferred astrophysical parameters flattens out.

We compare the two scenarios by computing the continuum reconstruction error $\boldsymbol{\delta}$ as defined in Eq.~(\ref{eq:reconstr_err}) with respect to the pure dimensionality-reduced continuum $\boldsymbol{s}_\mathrm{DR}$ as well as the inferred continuum $\boldsymbol{s}_\mathrm{inf}$. Note that we define the inferred continuum as the median of all model continua corresponding to the $\boldsymbol{\eta} = (s_\mathrm{norm}, \boldsymbol{\xi})$-samples of the full HMC chain. In order to distinguish the constraining power in the Lyman-$\alpha$ forest and damping wing region from that in the smooth emission line region of the spectrum, we separately compute the mean $\langle \boldsymbol{\delta} \rangle$ and the standard deviation $\sigma(\boldsymbol{\delta})$ of the reconstruction error with respect to the spectral pixels in the blue ($1185\,\text{\AA} \leq \lambda \leq 1260\,\text{\AA}$) and the red part ($1260\,\text{\AA} < \lambda \leq 2000\,\text{\AA}$) of the spectrum. We chose these ranges in agreement with those in Figures~3 and 4 in \citet{hennawi2024}.

We show the median values of these quantities with respect to the ensemble of $100$ mock objects as a function of latent dimension $n_\mathrm{latent}$ in Figure~\ref{fig:lat_dim_reconstr_err}. The behavior of the inferred and the dimensionality-reduced continua excellently agrees on the red side of the spectrum. Both the bias $\langle \boldsymbol{\delta} \rangle$ and the standard deviation $\sigma(\boldsymbol{\delta})$ decrease with increasing latent dimension due to the fact that more flexible models can always provide a closer fit to the true shape of the continuum \citep[see also Figures~3 and 4 in][]{hennawi2024}. As also observed in \citet{hennawi2024}, this curve does not flatten out before $n_\mathrm{latent} = 10$, and by naive expectation, the same should hold true on the blue side of the spectrum. Indeed, this is the case for the purely dimensionality-reduced spectrum. We observe the opposite trend, however, for the reconstruction errors of the inferred continuum whose standard deviation reaches a shallow minimum around $n_\mathrm{latent} \sim 5$ while the bias~$\langle \boldsymbol{\delta} \rangle$ oscillates between $-0.02$ and $0.0$ without a clear dependence on the value of $n_\mathrm{latent}$. Most particularly, $\sigma(\boldsymbol{\delta})$ diverges from the curve of the dimensionality-reduced continuum at around $n_\mathrm{latent} \gtrsim 5$ and starts to grow at an increasing rate.

We thus conclude that additional PCA vectors do keep improving the formal fit to the continuum, but they do not yield additional constraining power because they do not encode additional 
physical information about the blue side of the spectrum that is absorbed by the Lyman-$\alpha$ forest or the IGM damping wing. As a result,
the variation of the reconstruction error of the inferred continuum starts to increase instead of continuing to decrease for higher-dimensional PCA models. Choosing a model that is too flexible
is therefore not desirable for reconstructing the continuum accurately from a spectrum with a Lyman-$\alpha$ damping wing imprint. This insight, along with the weak dependence of inference precision $\Delta_{68\,\%}(\langle x_\mathrm{HI} \rangle)$ and $\Delta_{68\,\%}(\log_{10} t_\mathrm{Q}/\mathrm{yr})$ on latent dimension, lead us to the conclusion that $n_\mathrm{latent} = 5$ is an adequate choice for our PCA continuum model.
Note, however, that this value depends on the spectral coverage which was here analyzed up to $2000\,\text{\AA}$. Spectra covering additional wavelength ranges redward of $2000\,\text{\AA}$ would likely require PCA models of a higher dimensionality, but the general behavior found in this section would hold.

\subsubsection{Wavelength coverage}
\label{sec:wave_cut}

The smooth emission lines on the red side of a quasar spectrum are correlated with the shape of the continuum blueward of the Lyman-$\alpha$ line \citep[see Figure~7 in][]{hennawi2024}. How well we can reconstruct the continuum in this region is going to be degenerate with the astrophysical parameters $\langle x_\mathrm{HI} \rangle$ and $t_\mathrm{Q}$ governing the IGM transmission field. Gaining as much information as possible about the red-side shape of the continuum by covering more emission lines blueward of the Lyman-$\alpha$ line can thus enhance overall inference precision on the astrophysical parameters of interest.

Our fiducial wavelength grid 
covers the rest-frame wavelength range between $1185 \,\text{\AA}$ and $2000\,\text{\AA}$. In this section, we investigate how much information is contained in the outer regions on the red side of the spectrum, informing us about the importance of covering those regions when collecting new spectra for damping wing analysis.

Covering the inevitably required Lyman-$\alpha$ region with a given spectograph will automatically provide us with coverage out to some wavelength redward of the Lyman-$\alpha$ line. Getting redder coverage out to $2000\,\text{\AA}$ in the rest-frame would likely incur more observational cost to observe multiple setups (depending on the spectrograph) and as such we aim to quantify in this section the precision gains from higher red-side wavelength coverage.

The two upper panels of Figure~\ref{fig:mix_spec} show the example spectrum already considered in Figure~\ref{fig:lat_dim_spec}, cut off at $\lambda_\mathrm{cut} = 1300 \,\text{\AA}$, as well as the full spectrum with $\lambda_\mathrm{cut} = 2000 \,\text{\AA}$. The uncertainties of the inferred continuum are notably higher when cutting off the spectrum at $\lambda_\mathrm{cut} = 1300 \,\text{\AA}$. The same is the case for the inferred astrophysical parameters $\langle x_\mathrm{HI} \rangle$ and $t_\mathrm{Q}$ whose marginal posterior distributions are depicted in the second panel of Figure~\ref{fig:mix_contour}: 
inference precision significantly deteriorates for the shorter wavelength coverage of $\lambda_\mathrm{cut} = 1300 \,\text{\AA}$ which excludes the S IV emission line around $1397 \,\text{\AA}$ and the \ion{O}{IV}] line at $1402 \,\text{\AA}$, as well as the \ion{C}{IV} line around $1549 \,\text{\AA}$. Particularly the \ion{C}{IV} 
emission line strength and shape correlates strongly with the shape of the continuum in the Lyman-$\alpha$ region where we need to reconstruct the IGM damping wing imprint
\citep[again see Figure~7 in][]{hennawi2024}. 
An intermediate cut-off at $\lambda_\mathrm{cut} = 1600 \,\text{\AA}$ which includes all these lines produces posteriors whose shape is almost identical to those of the full spectrum (i.e., $\lambda_\mathrm{cut} = 2000 \,\text{\AA}$).

The same trends hold for the full set of $100$ mock quasar spectra. Between covering the full wavelength grid out to $\lambda_\mathrm{cut} = 2000 \,\text{\AA}$ and a cut-off wavelength of $\lambda_\mathrm{cut} = 1600 \,\text{\AA}$, the posterior widths largely remain constant and gradually start increasing for shorter cut-off wavelengths, as seen in the upper right panels of Figure~\ref{fig:mix_precision}. 
The same is the case for the subset of spectra from the fiducial region of parameter space. 
Therefore, major emission lines (especially the \ion{C}{IV} line) 
contain important information that significantly enhances the overall inference precision. The correlation with the weaker lines in the outer part of the spectrum, on the other hand, is much less pronounced. We can therefore conclude that inference quality does not suffer notably if wavelengths in the range between $1600 \,\text{\AA} \leq \lambda \leq 2000 \,\text{\AA}$ are missing due to observational constraints or cost (or if they are excluded to break down the computational complexity). 
Our analysis does not exclude, however, that important emission lines beyond $\lambda_\mathrm{cut} = 2000 \,\text{\AA}$, such as \ion{Mg}{II}, or even H-$\beta$ or H-$\alpha$ that are within the reach of JWST instruments, 
could further enhance inference precision. On the contrary, given the strong impact of the \ion{C}{IV} emission line, it is not unlikely 
that covering these lines would further improve the overall constraints. The challenge in this regard is constructing an adequately large PCA training set for our continuum DR model. We postpone this task to future work.

\subsubsection{Signal to Noise Ratio}
\label{sec:snr}

Required exposure times and thus the cost of future observations
crucially depends on the desired signal-to-noise ratio $\mathrm{S}/\mathrm{N}$ of the spectrum. Potential gains in inference precision from 
obtaining higher-$\mathrm{S}/\mathrm{N}$ spectra are thus highly informative about the most efficient setting for upcoming observations. Throughout this work, we define $\mathrm{S}/\mathrm{N}$ as the signal-to-noise ratio per $100\;\mathrm{km}/\mathrm{s}$ velocity interval, computed over the telluric absorption free observed frame wavelength range $11750-13300\,\text{\AA}$ (i.e., $1376-1557\,\text{\AA}$ in the rest-frame at $z_\mathrm{QSO} = 7.54$
).
Figure~\ref{fig:mix_spec} shows examples of inferred spectra for a very noisy ($\mathrm{S}/\mathrm{N} = 2$), a very high signal-to-noise ratio
($\mathrm{S}/\mathrm{N} = 20$), and our reference spectrum with $\mathrm{S}/\mathrm{N} = 10$. The random seed 
is the same for each realization and only the amplitude 
of the noise is adjusted. While the high signal-to-noise spectrum does not lead to significant improvements over the intermediate one, the low signal-to-noise spectrum has somewhat larger error bars in the inferred continuum. The shape of its marginalized posterior distribution is wider as well, as the contours in the third panel of Figure~\ref{fig:mix_contour} suggest.

The $100$ object ensemble considered in the bottom left panels in Figure~\ref{fig:mix_precision} confirms these trends: both neutral fraction and lifetime precision ($\Delta_{68\,\%}(\langle x_\mathrm{HI} \rangle)$ and $\Delta_{68\,\%}(\log_{10} t_\mathrm{Q}/\mathrm{yr})$) do not change significantly between $10 \lesssim \mathrm{S}/\mathrm{N} \lesssim 20$ and start increasing notably at smaller signal-to-noise ratios. The same trends are apparent for the subset of fiducial samples. We therefore conclude that spectra with signal-to-noise ratios of $\mathrm{S}/\mathrm{N} \sim 10$ are sufficient for a precise parameter inference.

\subsubsection{Spectral resolution}
\label{sec:fwhm}

Spectral resolution is one of the most important factors to consider for both observational and computational reasons. Modern-day instruments such as X-Shooter are able to collect spectra with resolutions of $R \sim 10\,000$ even for the faintest quasars that can be of interest for damping wing analysis.

Resolving the Lyman-$\alpha$ forest with such high resolutions might be beneficial since within the noise, each transmission spike constitutes a lower limit for the continuum at the corresponding wavelength pixel. In other words, the continuum should never fall below the observed data as the IGM transmission values can never exceed unity. Furthermore, resolving the proximity zone with additional resolution elements may further increase constraining power. On the other hand, higher resolutions hold little to no gains in the region of the spectrum redward of the Lyman-$\alpha$ line where the flux is a smooth function of wavelength. Hence, we divide the wavelength grid at $\lambda \simeq 1218\,\text{\AA}$ into a low-resolution red part and a blue part where we vary resolution between low values comparable to the red-side resolution all the way up to X-Shooter-like high-resolution values.

Yet, matrix inversion and therefore the cost of a likelihood evaluation scale cubically with the number of spectral pixels, and even after introducing the aforementioned hybrid wavelength grid, we end up with up to $O(10^4)$ spectral pixels, making the repeated inversion of the corresponding covariance matrices for the likelihood computations in the inference process notably more expensive. Besides the observational cost of collecting such high-resolution spectra, it is hence also relevant from a computational point of view to quantify the gains of analyzing spectra at high spectral resolution.

We remind the reader that we quantify spectral resolution in terms of the $\mathrm{FWHM}$ 
of the LSF. When varying $\mathrm{FWHM}$, we proportionally adjust the velocity pixel scale, keeping the spectral sampling factor fixed to $2$. Note also that we always work with a constant signal-to-noise ratio of $\mathrm{S}/\mathrm{N} = 10$ \textit{per} $100\,\mathrm{km}/\mathrm{s}$ velocity interval. This assures that we do not erroneously attribute signal-to-noise trends to spectral resolution.

Our technical implementation comes with the minor caveat that each mock realization of the spectrum at a different resolution has its unique 
noise draw. Though this acts as an additional source of stochasticity, we find that after averaging over all 100 mock spectra the effects are small enough to deduce an overall trend. 
The sample size of our $16$ fiducial spectra, however, is not sufficient to average out the statistical fluctuations due to the different noise draws. To gain a statistically meaningful picture in the fiducial region of $(\langle x_\mathrm{HI} \rangle, t_\mathrm{Q})$-parameter space, we instead fix the astrophysical parameters to the fiducial values of $\langle x_\mathrm{HI} \rangle = 0.5$ and $t_\mathrm{Q} = 10^6\;\mathrm{yr}$ and simulate a new set of $100$ mock spectra which only differ by the draws of the continuum and of the lognormal Lyman-$\alpha$ forest fluctuations.

Strikingly, the effects of changing spectral resolution are smaller than one might expect. Figure~\ref{fig:mix_spec} depicts the inferred continuum of our example spectrum for the reference resolution of $\mathrm{FWHM} = 100\;\mathrm{km}/\mathrm{s}$, a fine resolution of $\mathrm{FWHM} = 30\;\mathrm{km}/\mathrm{s}$, and a coarse value of $\mathrm{FWHM} = 1000\;\mathrm{km}/\mathrm{s}$ whose pixel scale of $500\;\mathrm{km}/\mathrm{s}$
coincides with that on the red side of the spectrum. The inferred continua look remarkably similar; even for the coarsest resolution, the general shape of the continuum is well captured. The corresponding posteriors in Figure~\ref{fig:mix_contour} are somewhat less informative but still largely retain their original shape even for $\mathrm{FWHM} = 1000\;\mathrm{km}/\mathrm{s}$.

We find similarly weak trends for the full ensemble of $100$ objects. The width of the posteriors in Figure~\ref{fig:mix_precision} only changes very slightly with resolution. In particular, the difference between the median widths of the $\langle x_\mathrm{HI} \rangle$ posteriors from the highest- and lowest-resolution runs is only $\sim 4\,\%$, and $\sim 0.03\,\mathrm{dex}$ for $\log_{10} t_\mathrm{Q}/\mathrm{yr}$.
This shows that spectral resolution can be lowered drastically almost without any notable impact on inference precision. The reason for this appears to be that resolving proximity zone absorption lines is secondary for a precise inference of the astrophysical parameters $\langle x_\mathrm{HI} \rangle$ and $t_\mathrm{Q}$. The main constraining power seems to come from the smooth IGM damping wing signature along with the size and shape of the quasar proximity zone which are still determined to within reasonable accuracy even for the lowest-resolution spectra (cf. Figure~\ref{fig:mix_spec}).
Merely
resolutions that resolve all individual Ly-$\alpha$ absorption lines could possibly lead to improvements but might not be preferable from a computational point of view, given the computational cost of analyzing higher resolution spectra and the limited benefit with respect to constraining power.

Note that in the case of the full-simulation IGM transmission model, higher resolution \textit{does} lead to an improvement in inference precision. The inferred posteriors, however, are significantly overconfident. This is a result of our Gaussian approximation for the IGM transmission likelihood (Eq.~(\ref{eq:P_t_theta})) being less adequate at increasingly high resolutions \citep[see also Figure~11 in][]{hennawi2024}. We elaborate in detail on coverage tests and inference precision of the full-simulation model in Appendices~\ref{app:inf_tests} and \ref{app:param_scan_sims}. After correcting for the overconfidence, the dependence of inference precision on spectral resolution is similarly weak as in the case of the simple analytic IGM transmission model. Informed by these trends, we adopt the coarse red-side resolution ($\mathrm{FWHM} = 1000\;\mathrm{km}/\mathrm{s}$) across the entire wavelength grid, pending further improvements in our likelihood approximation that we leave for future work. 
Given that the need to apply big coverage corrections is currently obviating the gains from higher resolution spectra, this choice of resolution minimizes the computational and observational cost of the analysis.
It is also advantageous from a computational point of view as it reduces the total number of spectral pixels to $n_\lambda = 313$. When applied to real-world data from a higher-resolution spectrograph, we opt to achieve a comparable setting by rebinning the spectrum to the corresponding velocity pixel scale of $500\;\mathrm{km}/\mathrm{s}$. This is also the strategy we adopt in the following section to characterize the overall error budget of our inference pipeline as realistically as possible.

\subsubsection{The optimal hyper-parameter setting}

In summary, based on the preceding results, we adopt the following choices as the optimal hyper-parameter setting for the subsequent quantification of the full error budget of our inference pipeline: we choose a latent dimension of $n_\mathrm{latent} = 5$ for the PCA dimensionality reduction model (along with the amplitude degree-of-freedom resulting in a six-parameter continuum model), a rest-frame wavelength coverage out to $\lambda_\mathrm{cut} = 2000\,\text{\AA}$, an intermediate signal-to-noise ratio of $\mathrm{S}/\mathrm{N} = 10$ per $100\;\mathrm{km}/\mathrm{s}$ velocity interval, and a spectral resolution of $\mathrm{FWHM} = 100\;\mathrm{km}/\mathrm{s}$ rebinned to a velocity pixel scale of $500\;\mathrm{km}/\mathrm{s}$. We emphasize that even though the analysis above was based on the analytical IGM transmission model, we arrive at the same hyper-parameter choices when modelling IGM transmission with our full-simulation model. We discuss these results in Appendix~\ref{app:param_scan_sims}. As pointed out above, the only notable difference between the two IGM transmission models with regard to hyper-parameter variations is the deteriorating coverage behavior in the full-simulation model which is obviating the gains from high-resolution spectra and led us to conclude that we can adopt the coarse red-side resolution across the entire spectrum. To eliminate any remaining impact from non-optimal coverage behavior, we apply the marginal coverage correction procedure described in Section~\ref{sec:inf_tests} also to the mock ensemble considered in the following section.

\subsection{The Total Error Budget}
\label{sec:error_budget}

\begin{figure*}
    \centering
	\includegraphics[height=0.82\textwidth]{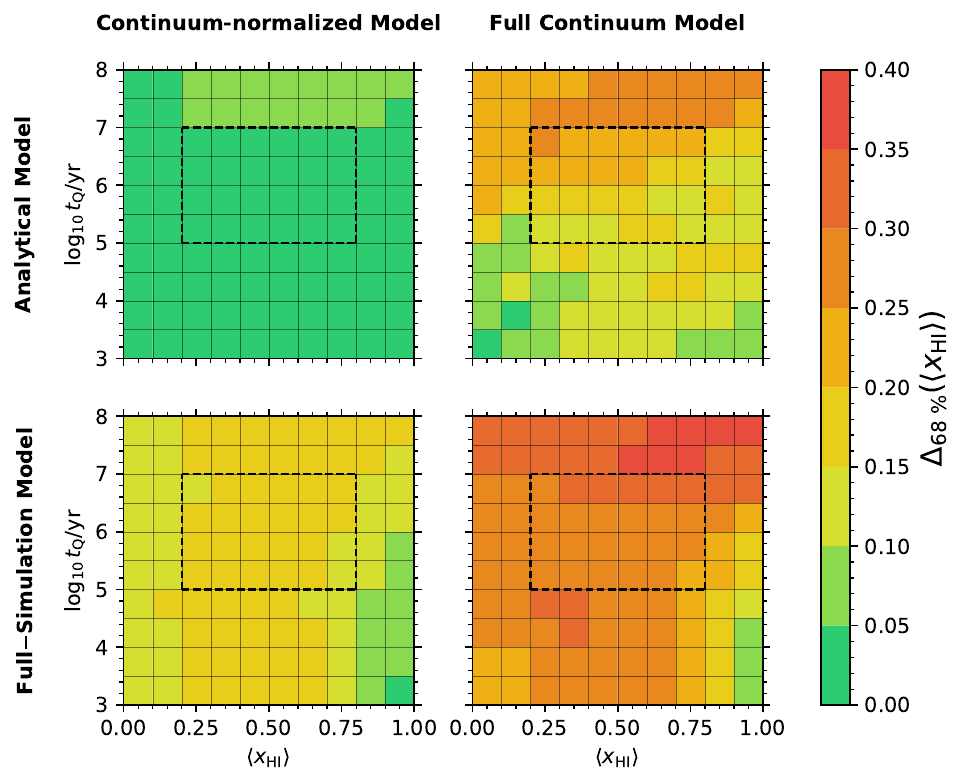}
    \caption{Inference precision with respect to IGM neutral fraction $\langle x_\mathrm{HI} \rangle$ as a function of $(\langle x_\mathrm{HI} \rangle, t_\mathrm{Q})$-parameter space, broken up into the individual model components. The two rows compare the two IGM transmission models, the columns show the effect of the continuum model. The dashed boxes enclose the fiducial region of parameter space, centered around $\langle x_\mathrm{HI} \rangle = 0.5$ and $t_\mathrm{Q} = 10^6\;\mathrm{yr}$. All precision values are computed from our $1000$ object ensemble, ran with the optimal hyper-parameter setting ($n_\mathrm{latent} = 5$, $\lambda_\mathrm{cut} = 2000\,\text{\AA}$, $\mathrm{S}/\mathrm{N} = 10$, and $\mathrm{FWHM} = 100\;\mathrm{km}/\mathrm{s}$ rebinned to $\mathrm{d}v_\mathrm{blue} = 500\;\mathrm{km}/\mathrm{s}$). Each pixel shows the average precision $\Delta_{68\,\%}(\langle x_\mathrm{HI} \rangle)$ with respect to the same $10$ continua. All precision values are based on reweighted marginal posterior PDFs, assuring that each mock ensemble passes a marginal coverage test with respect to the two astrophysical parameters $\langle x_\mathrm{HI} \rangle$ and $t_\mathrm{Q}$. Inference precision increases for objects with a stronger damping wing imprint. Both the continuum model and the stochasticity of reionization encoded in the full-simulation model add a similar amount of uncertainty, as quantified in Table \ref{tab:scatter_table}.}
    \label{fig:xHI_scatter}
\end{figure*}

\begin{figure*}
    \centering
	\includegraphics[height=0.82\textwidth]{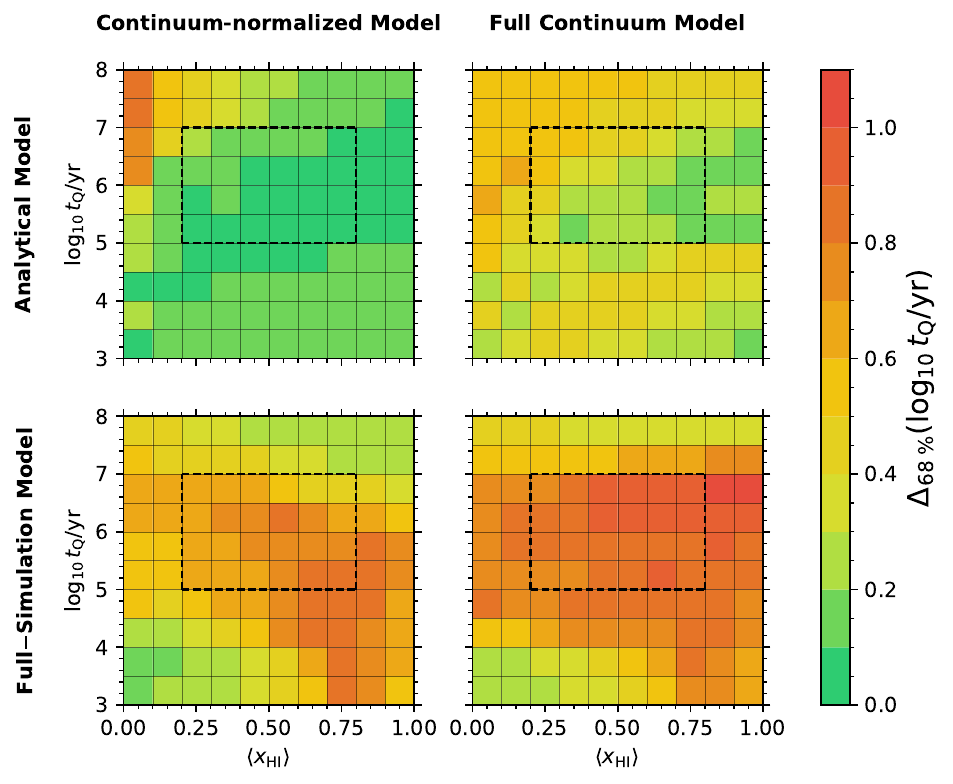}
    \caption{Like Figure~\ref{fig:xHI_scatter}, but for lifetime inference precision $\Delta_{68\,\%}(\log_{10} t_\mathrm{Q}/\mathrm{yr})$. For the analytical IGM transmission model, a stronger damping wing imprint implies higher lifetime precision, as seen before in the case of $\Delta_{68\,\%}(\langle x_\mathrm{HI} \rangle)$. The stochasticity of the full-simulation model erodes these trends. The longest-lived objects show extraordinarily high precision due to the thermal proximity effect.}
    \label{fig:tQ_scatter}
\end{figure*}

\begin{table*}
	\centering
	\caption{Median $\langle x_\mathrm{HI} \rangle$- and $t_\mathrm{Q}$-inference precision $\Delta_{68\,\%}(\langle x_\mathrm{HI} \rangle)$ and $\Delta_{68\,\%}(\log_{10} t_\mathrm{Q}/\mathrm{yr})$, broken up into the individual model components as seen in Figures~\ref{fig:xHI_scatter} and \ref{fig:tQ_scatter}. %
    The values listed are medians of all samples in the full (or fiducial) region of parameter space, given the corresponding model configuration. The parameter limits of the fiducial region are enclosed by dashed boxes in Figures~\ref{fig:xHI_scatter} and \ref{fig:tQ_scatter}. Uncertainties are estimated according to the respective $16\,\%$ and $84\,\%$ percentile values.}
	\label{tab:scatter_table}
	\begin{tabular}{rr|cc|cc} %
		\hline
		& & \multicolumn{2}{c}{$\Delta_{68\,\%}(\langle x_\mathrm{HI} \rangle)$} & \multicolumn{2}{c}{$\Delta_{68\,\%}(\log_{10} t_\mathrm{Q}/\mathrm{yr})$} \\
		\cmidrule(lr){3-4} \cmidrule(lr){5-6}
        & & Continuum-normalized & Full continuum model & Continuum-normalized & Full continuum model \\
		\hline \hline
		\multirow{2}{*}{All samples} & Analytical Model & $2.2_{-0.8}^{\hspace{.056em}+\hspace{.056em}2.5}\,\%$ & $15.2_{-\phantom{1}6.9}^{\hspace{.056em}+\hspace{.056em}12.8}\,\%$ & $0.12_{-0.06}^{\hspace{.056em}+\hspace{.056em}0.12}\;\mathrm{dex}$ & $0.36_{-0.18}^{\hspace{.056em}+\hspace{.056em}0.19}\;\mathrm{dex}$ \vspace{.5em} \\
		& Full-Simulation Model & $15.2_{-4.2}^{\hspace{.056em}+\hspace{.056em}3.0}\,\%$ & $28.0_{-8.8}^{\hspace{.056em}+\hspace{.056em}8.2}\,\%$ & $0.55_{-0.34}^{\hspace{.056em}+\hspace{.056em}0.35}\;\mathrm{dex}$ & $0.80_{-0.55}^{\hspace{.056em}+\hspace{.056em}0.22}\;\mathrm{dex}$ \\
        \hline
		\multirow{2}{*}{Fiducial samples} & Analytical Model & $2.4_{-0.7}^{\hspace{.056em}+\hspace{.056em}1.1}\,\%$ & $17.4_{-5.9}^{\hspace{.056em}+\hspace{.056em}9.3}\,\%$ & $0.08_{-0.03}^{\hspace{.056em}+\hspace{.056em}0.06}\;\mathrm{dex}$ & $0.28_{-0.11}^{\hspace{.056em}+\hspace{.056em}0.20}\;\mathrm{dex}$ \vspace{.5em} \\
		& Full-Simulation Model & $15.8_{-2.4}^{\hspace{.056em}+\hspace{.056em}2.4}\,\%$ & $28.5_{-6.5}^{\hspace{.056em}+\hspace{.056em}7.1}\,\%$ & $0.72_{-0.32}^{\hspace{.056em}+\hspace{.056em}0.19}\;\mathrm{dex}$ & $0.95_{-0.21}^{\hspace{.056em}+\hspace{.056em}0.09}\;\mathrm{dex}$ \\
		\hline
	\end{tabular}
    \label{tab:prec}
\end{table*}

All inference runs in the previous sections were obtained with the analytical IGM transmission model described in Section~\ref{sec:analytic}. Its deterministic relationship between the astrophysical parameters $\boldsymbol{\theta} = (\langle x_\mathrm{HI} \rangle, t_\mathrm{Q})$ and damping wing strength allowed us to gain a maximally clear picture of how inference precision depends on the hyper-parameters considered. In doing so, we neglected the fact that in realistic models of reionization such as the \texttt{21cmFAST} topologies of our full-simulation model, the size of reionization bubbles does not depend deterministically on the global IGM neutral fraction. In addition, by performing radiative transfer calculations, we account for the fact that fluctuations in the HI density field will impact the size of the quasar's ionization front which we previously approximated deterministically according to Eq.~(\ref{eq:R_ion}).

In this section, we account for these two sources of stochasticity by modelling IGM transmission according to our full-simulation model introduced in Section~\ref{sec:sims}. Re-running an identical set of mock objects, combined with analytical IGM transmission skewers, allows us to precisely quantify the impact of the aforementioned sources of stochasticity on the total error budget of our inference pipeline. By repeating the inference after dividing each spectrum by its underlying continuum, we are further able to quantify the contribution of the continuum reconstruction error on the final inference precision. In other words, we break down the total error budget of our inference scheme into the separate contributions from the continuum model and the two aforementioned additional sources of stochasticity encoded in our full-simulation model.

We further investigate how these contributions change depending on the exact location in $(\langle x_\mathrm{HI} \rangle, t_\mathrm{Q})$-parameter space. To that end, we define a linear $(10 \times 10)$-grid covering our full $(\langle x_\mathrm{HI} \rangle, \log_{10} t_\mathrm{Q}/\mathrm{yr})$-prior range, i.e., $0 \leq \langle x_\mathrm{HI} \rangle \leq 1$ and $3 \leq \log_{10} t_\mathrm{Q}/\mathrm{yr} \leq 8$. In addition, we draw $10$ distinct autofit continua and combine them with $10$ distinct IGM transmission skewers at each location of the $(\langle x_\mathrm{HI} \rangle, \log_{10} t_\mathrm{Q}/\mathrm{yr})$-grid. By considering the same $10$ continua at each point in astrophysical parameter space, we suppress additional stochasticity that would otherwise arise due to varying continuum draws. This results in an overall ensemble of $(10\times10)\times10 = 1000$ mock spectra.

Figure~\ref{fig:xHI_scatter} summarizes all aforementioned dependencies for $\langle x_\mathrm{HI} \rangle$-inference precision $\Delta_{68\,\%}(\langle x_\mathrm{HI} \rangle)$. The four different panels show inference precision in $(\langle x_\mathrm{HI} \rangle, \log_{10} t_\mathrm{Q}/\mathrm{yr})$-parameter space for the same $1000$ mock objects considered with/without continuum model and with analytic/full-simulation IGM transmission model, respectively.
Lifetime inference precision $\Delta_{68\,\%}(\log_{10} t_\mathrm{Q}/\mathrm{yr})$ is shown in Figure~\ref{fig:tQ_scatter}. Table \ref{tab:prec} lists the respective median precision values with respect to the full prior volume and the fiducial region of parameter space as defined in Section~\ref{sec:inf_prec} (enclosed by the dashed boxes in Figures~\ref{fig:xHI_scatter} and \ref{fig:tQ_scatter}).

Independently of the model configuration, we see significant variation as a function of the astrophysical parameter values. As an underlying general trend, we find an overall improvement in constraining power for short quasar lifetimes $t_\mathrm{Q}$, and, to a somewhat lesser extent, for higher neutral fractions $\langle x_\mathrm{HI} \rangle$. This behavior can be directly attributed to the enhanced damping wing strength of quasars in a neutral environment and short-lived quasars that have had less time to carve out a large ionized bubble in their immediate proximity. The trend is particularly pronounced for the analytic IGM transmission model where this dependence on $\langle x_\mathrm{HI} \rangle$ and $t_\mathrm{Q}$ is incorporated explicitly through Eqs.~(\ref{eq:R_ion}) and (\ref{eq:tau_DW}).

Secondly, we point out that inference precision can vary notably between the same continua combined with analytical or full-simulation IGM transmission skewers. We observe a significantly increased amount of scatter for the full-simulation model and overall deteriorating inference precision in most regions of parameter space. This is an immediate effect of the additional sources
of stochasticity due to the non-deterministic bubble-size distribution from reionization, and the size of the ionized bubble sourced by the quasar. 
As a result, in our full model, $\Delta_{68\,\%}(\langle x_\mathrm{HI} \rangle)$ can be as low as $3.0\,\%$ and as high as $42.2\,\%$ for individual objects. Likewise, $\Delta_{68\,\%}(\log_{10} t_\mathrm{Q}/\mathrm{yr})$ varies between $0.07\,\mathrm{dex}$ and $1.12\,\mathrm{dex}$. The full-parameter space medians 
of $\Delta_{68\,\%}(\langle x_\mathrm{HI} \rangle) = 28.0_{-8.8}^{\hspace{.056em}+\hspace{.056em}8.2}\,\%$ and $\Delta_{68\,\%}(\log_{10} t_\mathrm{Q}/\mathrm{yr}) = 0.80_{-0.55}^{\hspace{.056em}+\hspace{.056em}0.22}\;\mathrm{dex}$, respectively, 
are dominated to a similar extent by the errors related to continuum reconstruction and the stochasticity of the ionized bubble size distribution: for the most basic task of exclusively inferring the two astrophysical parameters $\langle x_\mathrm{HI} \rangle$ and $t_\mathrm{Q}$ from a continuum-normalized spectrum with deterministic IGM transmission model, we are able to recover $\langle x_\mathrm{HI} \rangle$ with only $2.2_{-0.8}^{\hspace{.056em}+\hspace{.056em}2.5}\,\%$ uncertainty as quantified in terms of $\Delta_{68\,\%}(\langle x_\mathrm{HI} \rangle)$. Modelling the 
quasar continuum and switching to the full-simulation IGM transmission model encoding the stochasticity of reionization and the quasar ionized bubble sizes
raises $\Delta_{68\,\%}(\langle x_\mathrm{HI} \rangle)$ to a total of $28.0_{-8.8}^{\hspace{.056em}+\hspace{.056em}8.2}\,\%$.We can see that the addition of continuum reconstruction error
and the full-simulation IGM transmission model contribute a roughly equal amount to the total error budget by turning them on separately: the continuum model raises $\Delta_{68\,\%}(\langle x_\mathrm{HI} \rangle)$ to $15.2_{-\phantom{1}6.9}^{\hspace{.056em}+\hspace{.056em}12.8}\,\%$, and the full-simulation IGM transmission model to the almost identical value of $15.2_{-4.2}^{\hspace{.056em}+\hspace{.056em}3.0}\,\%$ (see Table \ref{tab:prec}). Most notably, this implies that the intrinsic stochasticity of reionization and quasar ionized bubble sizes accounts for almost half of the total error budget. Eliminating this stochasticity from the inference process could lead to significant precision gains in future inference pipelines.

The overall picture for lifetime precision $\Delta_{68\,\%}(\log_{10} t_\mathrm{Q}/\mathrm{yr})$ is similar. We can infer $t_\mathrm{Q}$ from a continuum-normalized spectrum with analytical IGM transmission model with an uncertainty of $0.12_{-0.06}^{\hspace{.056em}+\hspace{.056em}0.12}\;\mathrm{dex}$
This increases to $0.36_{-0.18}^{\hspace{.056em}+\hspace{.056em}0.19}\;\mathrm{dex}$ when modelling the quasar continuum and keeping the IGM transmission model deterministic, and to $0.55_{-0.34}^{\hspace{.056em}+\hspace{.056em}0.35}\;\mathrm{dex}$ when switching to the full-simulation IGM transmission model but not performing continuum reconstruction,
adding up to a total of $0.80_{-0.55}^{\hspace{.056em}+\hspace{.056em}0.22}\;\mathrm{dex}$ when both models are turned on simultaneously. Thus, the intrinsic stochasticity of reionization and the quasar ionized bubble sizes even dominates over continuum reconstruction uncertainties in the total lifetime error budget. When comparing the full-simulation to the analytical IGM transmission model, an additional feature in $(\langle x_\mathrm{HI} \rangle, t_\mathrm{Q})$-parameter space is worthy of note: 
lifetime uncertainties of the oldest objects ($t_\mathrm{Q} \gtrsim 10^7\,\mathrm{yr}$) are significantly smaller than those of intermediate-lifetime ones. This opposes the general trend that we pointed out in Section~\ref{sec:inference} 
(i.e., the longer the lifetime, the larger the ionized region around the quasar and hence the weaker the damping wing signature, exacerbating the parameter inference). Notably, the lifetime uncertainties of parts of these objects are even smaller than those of their analytical counterparts. We attribute this improvement in precision to the thermal proximity effect of He II reionization 
\citep{Meiksin2010, Bolton2010, Bolton2012, Khrykin2016}: the photoelectric heating resulting from
the He II $\to$ He III photoionizations gives rise to a thermal proximity zone around the quasar with a significantly increased IGM temperature, causing a change in shape of the IGM transmission profile via the Gunn-Peterson optical depth formula. Eq.~(\ref{eq:tau_GP}) of our analytical IGM transmission model, on the other hand, does not account for this effect.
The size of the thermal proximity zone increases with quasar lifetime, facilitating the inference for the longest-lifetime objects. As a result, objects in the fiducial region of parameter space with weaker He II proximity effect suffer the most under the more realistic IGM transmission model based on cosmological simulations, as evident in a fiducial parameter space median lifetime precision of $0.95_{-0.21}^{\hspace{.056em}+\hspace{.056em}0.09}\;\mathrm{dex}$, with $0.72_{-0.32}^{\hspace{.056em}+\hspace{.056em}0.19}\;\mathrm{dex}$ caused by the stochasticity of reionization alone.

\section{Conclusions}
\label{sec:conclusions}

In this work we explored 
a pipeline to infer the global volume-averaged IGM neutral fraction and the quasar lifetime using the damping wing signature of high-redshift quasars. We determined the optimal hyper-parameter setting and rigorously quantified the associated error budget. The inference pipeline has been introduced in \citet{hennawi2024} and consists of a parametric (PCA) model for the quasar continuum, as well as a realistic forward model for IGM transmission based on skewers extracted from hydrodynamical simulations and a semi-numerical reionization topology combined with 1d ionizing radiative transfer. To isolate the impact of the stochasticity of reionization as well as the quasar ionized bubble sizes, we also introduced a deterministic analytical IGM transmission model.
We tested the pipeline on a fixed set of mock observational spectra with heteroscadastic observational noise, varying the latent dimension of our PCA continuum model, the wavelength coverage in the red part of the spectrum, the signal-to-noise ratio as well as spectral resolution. We found that:
\begin{enumerate}
    \item Latent dimensions of $n_\mathrm{latent} \sim 4 - 5$ are sufficient to capture all information that is intrinsically related to the IGM damping wing imprint. This conclusion differs from the pure dimensionality reduction task where \citet{hennawi2024} found the continuum reconstruction error to decrease monotonically as a function of $n_\mathrm{latent}$. Our results suggest that PCA vectors beyond $n_\mathrm{latent} > 5$ do not encode information about the damping wing signature and thus do not yield additional constraining power.
    \item Longer wavelength coverage in the red part of the spectrum improves inference precision, especially including major emission lines such as the \ion{Si}{IV} and the \ion{O}{IV}] line around $1400\,\text{\AA}$, as well as the \ion{C}{IV} line around $1549\,\text{\AA}$. However, if this information is not available, even the Lyman-$\alpha$ damping wing signature alone can yield non-trivial constraints. The trends suggest potential additional gains from covering major emission lines beyond our maximum wavelength coverage of $2000\,\text{\AA}$, such as \ion{Mg}{II} as well as H-$\beta$ or H-$\alpha$.
    \item A higher signal-to-noise ratio leads to higher precision, with $\mathrm{S}/\mathrm{N} \gtrsim 10$ per $100 \, \mathrm{km}/\mathrm{s}$ velocity stretch
    being desirable for sufficiently precise results. We note for the full-simulation IGM transmission model that higher $\mathrm{S}/\mathrm{N}$ comes at the price of decreasing coverage probability as our likelihood approximation fails to capture non-Gaussian information that is otherwise gaussianized by the noise.
    \item Given the current implementation of the inference pipeline, spectral resolution does not have a notable impact on inference precision. All gains that are achieved in the case of the full-simulation IGM transmission model are obviated by the coverage corrections we have to apply to the increasingly overconfident posteriors. We conclude that, pending improvements in our pipeline, we can reduce computational as well as observational cost by using the coarse red-side resolution of $\mathrm{d}v = 500 \, \mathrm{km}/\mathrm{s}$ for the entire spectrum.

\end{enumerate}
Given the optimal hyper-parameter setting, we quantified the individual uncertainty contributions of the continuum model as well as the stochasticity of the 
ionized bubble size distribution as well as the quasar ionized bubble sizes as a function of astrophysical parameter space. We concluded that:
\begin{enumerate}
    \item Inference precision varies strongly across $(\langle x_\mathrm{HI} \rangle, t_\mathrm{Q})$-parameter space, with the highest precision achieved for objects with a strong damping wing imprint (i.e., young quasars in a largely neutral environment). The observationally most relevant parameter values lie in the intermediate precision range. Inferring very long quasar lifetimes ($t_\mathrm{Q} \gtrsim 10^7\,\mathrm{yr}$) is facilitated by the thermal proximity effect.

    \item Using our most sophisticated model setting (continuum model + full-simulation IGM transmission model), we can infer $\langle x_\mathrm{HI} \rangle$ to a median uncertainty of $28.0_{-8.8}^{\hspace{.056em}+\hspace{.056em}8.2}\,\%$ across the full parameter space covered by our simulation grid (i.e., $0 \leq \langle x_\mathrm{HI} \rangle \leq 1$ and $3 \leq \log_{10} t_\mathrm{Q}/\mathrm{yr} \leq 8$), and to $28.5_{-6.5}^{\hspace{.056em}+\hspace{.056em}7.1}\,\%$ in the fiducial region between $0.25 \leq \langle x_\mathrm{HI} \rangle \leq 0.75$ and $5.25 \leq \log_{10} t_\mathrm{Q}/\mathrm{yr} \leq 6.75$. We can constrain quasar lifetimes to $0.80_{-0.55}^{\hspace{.056em}+\hspace{.056em}0.22}\;\mathrm{dex}$ and $0.95_{-0.21}^{\hspace{.056em}+\hspace{.056em}0.09}\;\mathrm{dex}$ in the full and fiducial region of parameter space, respectively.

    \item Across full parameter space, the (PCA-based) quasar continuum model contributes $15.2_{-\phantom{1}6.9}^{\hspace{.056em}+\hspace{.056em}12.8}\,\%$ to the total error budget of $\langle x_\mathrm{HI} \rangle$, and a roughly equal amount of $15.2_{-4.2}^{\hspace{.056em}+\hspace{.056em}3.0}\,\%$ is due to the stochasticity of reionization and quasar ionized bubble sizes. Across the fiducial region of parameter space, we find contributions of $17.4_{-5.9}^{\hspace{.056em}+\hspace{.056em}9.3}\,\%$ and $15.8_{-2.4}^{\hspace{.056em}+\hspace{.056em}2.4}\,\%$, respectively. Concerning lifetimes, the contributions of continuum model and stochasticity of reionization are $0.36_{-0.18}^{\hspace{.056em}+\hspace{.056em}0.19}\;\mathrm{dex}$ and $0.55_{-0.34}^{\hspace{.056em}+\hspace{.056em}0.35}\;\mathrm{dex}$, respectively, across the full parameter space, and $0.28_{-0.11}^{\hspace{.056em}+\hspace{.056em}0.20}\;\mathrm{dex}$ and $0.72_{-0.32}^{\hspace{.056em}+\hspace{.056em}0.19}\;\mathrm{dex}$ in the fiducial region, showing that the total error budget is largely dominated by the stochasticity of reionization and quasar ionized bubble sizes.
\end{enumerate}

Our results pave the way for the application of our inference machinery to real observational spectra to obtain robust constraints on the timing of reionization as well as the growth of supermassive black holes.
Combining the individual constraints of $O(10-100)$ independent objects at a similar redshift \citep[as are soon to be detected by the Euclid wide-field survey,][]{Euclid2019}
will scale down the aforementioned uncertainties by a factor of $\sqrt{10}-\sqrt{100}$, allowing us to place $\lesssim 5\,\%$-constraints on the evolution of the IGM neutral fraction as a function of redshift.

\section*{Acknowledgements}

We thank John Tamanas for contributions to an earlier version of the inference code base. 
We acknowledge helpful conversations with the ENIGMA group at UC Santa Barbara and Leiden University.
This work made use of \texttt{NumPy} \citep{Numpy2020},  \texttt{SciPy} \citep{Scipy2020}, \texttt{JAX} \citep{Jax2018}, \texttt{NumPyro} \citep{NumPyro2019a, NumPyro2019b}, \texttt{sklearn} \citep{sklearn2011}, \texttt{Astropy} \citep{Astropy2013, Astropy2018, Astropy2022}, \texttt{SkyCalc\_ipy} \citep{SkyCalc_ipy2021}, \texttt{h5py} \citep{h5py2013}, \texttt{Matplotlib} \citep{Matplotlib2007}, \texttt{corner.py} \citep{Corner2016}, and \texttt{IPython} \citep{IPython2007}.
This work was performed using the compute resources from the Academic Leiden Interdisciplinary Cluster Environment (ALICE) provided by Leiden University.
TK and JFH acknowledge support from the European Research Council (ERC) under the European Union’s Horizon 2020 research and innovation program (grant agreement No 885301). JFH acknowledges support from NSF grant No. 2307180.

\section*{Data Availability}

The derived data generated in this research will be shared on reasonable requests to the corresponding author.

\bibliographystyle{mnras}
\bibliography{allrefs} %

\appendix

\section{Correction to the Analytical Damping Wing Formula}
\label{app:mortlock}

\begin{figure*}
    \centering
	\includegraphics[width=\textwidth]{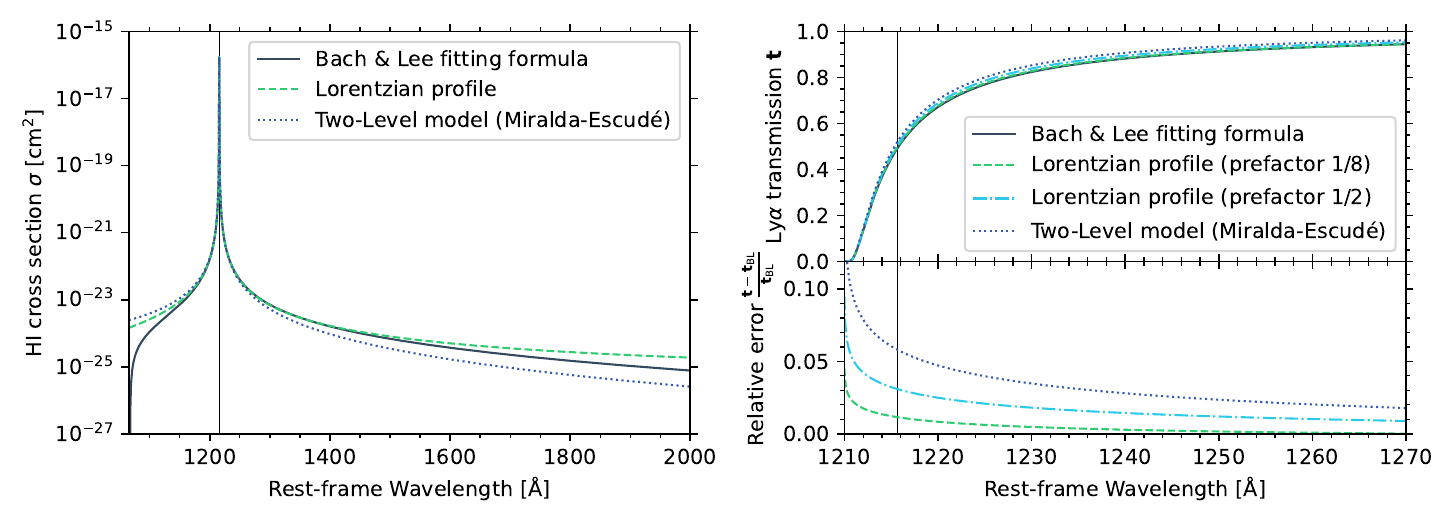}
    \caption{Approximations of the Lyman-$\alpha$ cross section of neutral hydrogen (left) and associated Lyman-$\alpha$ transmission field $\boldsymbol{t}$ in the damping wing region of the spectrum (right). The \citet{Bach2015} profile is an empirical fit to the full quantum-mechanical cross section; the Lorentzian profile (Eqs.~(\ref{eq:cross_section}) - (\ref{eq:integral})) and the two-level model \citep[Eq.~(\ref{eq:integral_2L}),][]{Miralda-Escude1998} are two simplified analytical approximations thereof. The transmission fields are calculated by integrating Eq.~(\ref{eq:tau_int}) directly using the \citet{Bach2015} fitting formula, or substituting Eqs.~(\ref{eq:integral}), (\ref{eq:integral_wrong}) and (\ref{eq:integral_2L}), respectively, into Eq.~(\ref{eq:tau_DW}) assuming $\langle x_\mathrm{HI} \rangle = 1$, $z_\mathrm{QSO} = 7.54$, $z_\mathrm{ion} = 7.5$ (corresponding to a reionization front at $R_\mathrm{ion} = 1.5\,\mathrm{pMpc}$) and $z_\mathrm{end} = 6.0$. The lower right panel shows the relative error of the approximate transmission profiles with respect to that obtained with the \citet{Bach2015} fitting formula. The vertical line highlights the Lyman-$\alpha$ resonance $\lambda_\alpha = 1216.78\,\text{\AA}$.}
    \label{fig:dw_profiles}
\end{figure*}

In Section~\ref{sec:analytic} we introduced an analytical expression (Eq.~(\ref{eq:tau_DW})) governing the damping wing optical depth. Our derivation closely follows that in \citet{Mortlock2016} but differs in a prefactor of $1/8$ in the second term of Eq.~(\ref{eq:integral}) instead of $1/2$ in Eq.~(40) in \citet{Mortlock2016}. This factor arises when solving the dimensionless integral
\begin{align}
    I(x) = \int \frac{x^{1/2}}{4(1-x)^2} \;\mathrm{d}x
    &= \frac{x^{1/2}}{4(1-x)} + \frac{1}{8} \ln\left(\frac{1-x^{1/2}}{1+x^{1/2}}\right) \\
    &\neq \frac{x^{1/2}}{4(1-x)} + \frac{1}{2} \ln\left(\frac{1-x^{1/2}}{1+x^{1/2}}\right).
\label{eq:integral_wrong}
\end{align}
Parts of our results (i.e., Section~\ref{sec:param_scan} and Appendices~\ref{app:inf_tests} and \ref{app:param_scan_sims}), were obtained with the incorrect prefactor of $1/2$. We investigate the effect of the correction in Figure~\ref{fig:dw_profiles} where we show the relative deviation of both $\tau_\mathrm{DW}$ prescriptions with respect to to that obtained from the full quantum mechanical Kramers-Heisenberg cross section of neutral hydrogen. We treat as the ground truth for this an empirical fitting formula introduced by \citet{Bach2015} that was designed as a simple but maximally accurate analytical approximation of the quantum mechanical cross section. For completeness, we also compare to the commonly used two-level model introduced by \citet{Miralda-Escude1998} \citep[originally derived in][]{Peebles1993} where
\begin{equation}
\label{eq:integral_2L}
    I(x) = \int \frac{x^\frac{9}{2}}{4(1-x)^2}.
\end{equation}
For each model, we determine the damping wing optical depth according to Eq.~(\ref{eq:tau_DW}) and compute the relative error with respect to the optical depth obtained by directly integrating Eq.~(\ref{eq:tau_int}) with the \citet{Bach2015} fitting formula. We find that the Lorentzian assumption for the cross section overestimates damping wing absorption by maximally $\sim 2\,\%$ over most parts of the relevant wavelength range. Working with the incorrect prefactor instead leads to an error of $\sim 1-2\,\%$, increasing to maximally $\sim 5\,\%$. The error of the \citet{Miralda-Escude1998} model is roughly twice as large across the entire wavelength range. In Section~\ref{sec:param_scan} and Appendix~\ref{app:param_scan_sims}, we were interested in the \textit{scaling} of inference precision with respect to various hyper-parameters rather than exact quantitative constraints. Due to this fact, and given the small error of the Lorentzian model with the incorrect prefactor of $1/2$ compared to the still widely used \citet{Miralda-Escude1998} model \citep[e.g.][]{Durovcikova2020, Lidz2021, Curtis-Lake2023, Heintz2023, Umeda2023}, we decided not to repeat all computations to save resources by avoiding significant computational overhead. The quantitative breakdown of the total error budget in Section~\ref{sec:error_budget} was obtained with the correct prefactor of $1/8$ and therefore remains unaffected.

\section{Inference Tests}
\label{app:inf_tests}

\begin{figure*}
    \centering
	\includegraphics[width=\textwidth]{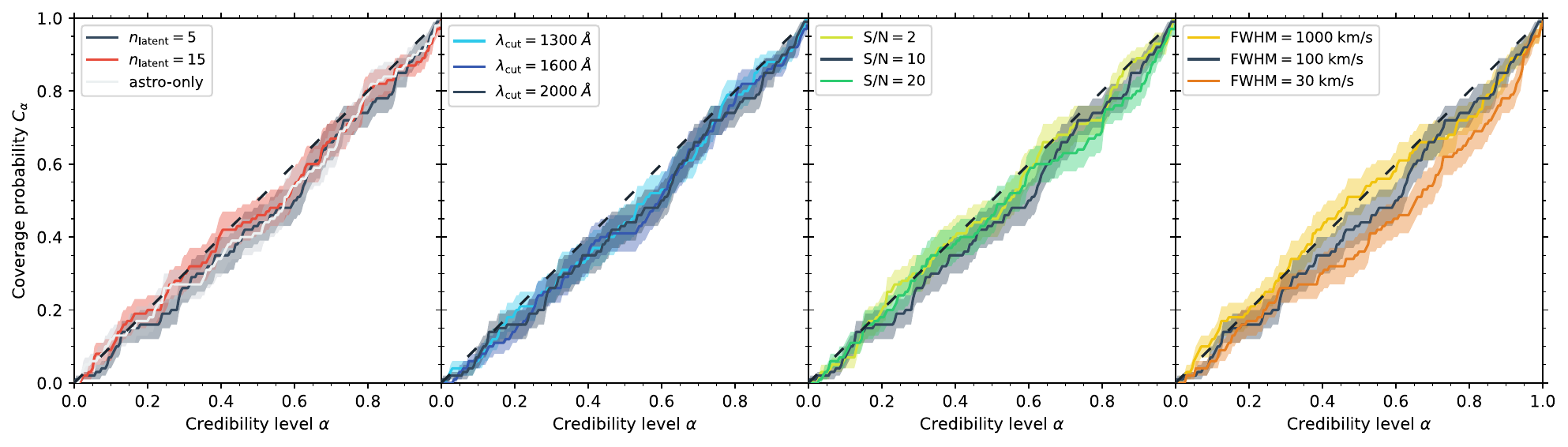}
    \caption{Marginal inference test results for the runs considered in Section~\ref{sec:param_scan}. In each panel, we show the expected coverage probability $C_\alpha$ as a function of the credibility level $\alpha$. Uncertainties correspond to the $16$-th and $84$-th percentile of a Binomial distribution $B(N, C_\alpha)$, where $N$ is the number of objects in our mock sample. Models with perfect coverage follow the black dashed line; the region above it indicates underconfidence, the region below overconfidence. Whenever applicable, we marginalized over all latent parameters related to the shape of the continuum before fitting a Gaussian mixture model (GMM) to the resulting marginal $(\langle x_\mathrm{HI} \rangle, \log_{10} t_\mathrm{Q}/\mathrm{yr})$-posterior density to compute the coverage probability. We vary latent dimension $n_\mathrm{latent}$, red-side wavelength coverage $\lambda_\mathrm{cut}$, signal-to-noise ratio $\mathrm{S}/\mathrm{N}$, and spectral resolution $\mathrm{FWHM}$, in the four panels from left to right. All inference test results were obtained using the analytical IGM transmission model. Each panel contains three representative choices of the respective parameter. Our reference run with $n_\mathrm{latent} = 5$, $\lambda_\mathrm{cut} = 2000\,\text{\AA}$, $\mathrm{S}/\mathrm{N} = 10$ and $\mathrm{FWHM} = 100\;\mathrm{km}/\mathrm{s}$) is depicted in black in each panel. The latent dimension plot also shows the theoretically optimal posteriors as obtained from the run with continuum-normalized spectra labelled as 'astro-only'.}
    \label{fig:coverage_analytic}
\end{figure*}

\begin{figure*}
    \centering
	\includegraphics[width=\textwidth]{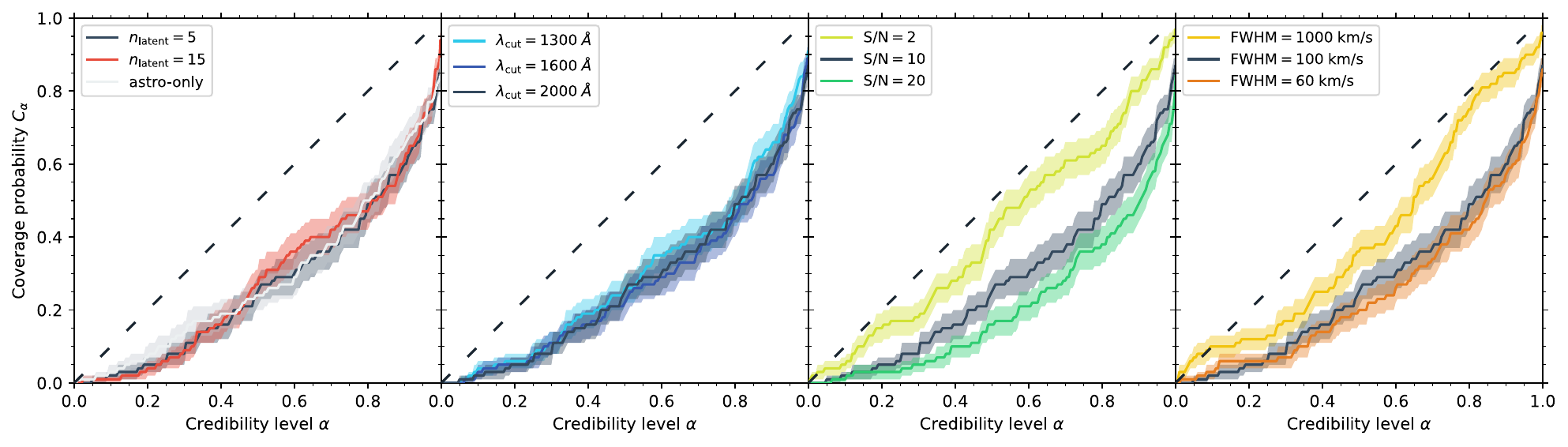}
    \caption{Like Figure~\ref{fig:coverage_analytic}, but for the full-simulation IGM transmission model.}
    \label{fig:coverage_sims}
\end{figure*}

In Section~\ref{sec:inf_tests} we introduced the concept of coverage tests to ensure 
that the inferred distribution of an ensemble of mock objects is a faithful representation of the true distribution. In Figure~\ref{fig:coverage_analytic}, we present the inference test results for the hyper-parameter configurations considered in Section~\ref{sec:param_scan} with the analytical IGM transmission model. 
For clarity, we show in each panel the expected coverage probability $C_\alpha$ of the marginal $(\langle x_\mathrm{HI} \rangle, \log_{10} t_\mathrm{Q}/\mathrm{yr})$-posteriors
for three representative choices of a given hyper-parameter. We observe perfect coverage to within $1\sigma$ almost independently of the chosen parameter values, where we assigned error bars based on a Binomial distribution $B(N, C_\alpha)$. This owes to the fact that $C_\alpha$ is the fraction of mocks which are fall inside the $\alpha$-th credibility region, and we estimate $C_\alpha$ by checking this for each of our $N=100$ mocks. We find that only spectral resolution has a slight impact on coverage probability. Our Gaussian likelihood approximation (Eq.~(\ref{eq:likelihood})) is thus remarkably accurate when IGM transmission is modelled according to the analytical formalism introduced in Section~\ref{sec:analytic}. 
Decreasing resolution changes the coverage probability from slightly overconfident to slightly underconfident for the lowest resolution spectra with $\mathrm{FWHM} = 1000\,\mathrm{km}/\mathrm{s}$, owing to the fact that less resolved spectral features (particularly in the Lyman-$\alpha$ forest) gaussianize the likelihood of the spectrum according to the central limit theorem. Accordingly, Eq.~(\ref{eq:likelihood}) becomes increasingly appropriate for lower resolution data. The extent as to which higher resolution models are overconfident is so small that the raw posteriors are almost identical to the reweighted ones that we obtain by applying the coverage correction procedure described in Section~\ref{sec:inf_tests}. Nonetheless, for consistency, we choose to work with the reweighted posteriors for all our results shown in Sections~\ref{sec:param_scan} and \ref{sec:error_budget} and Appendix~\ref{app:param_scan_sims}.

Our full-simulation IGM transmission model, on the other hand, treats more
complex physical processes which give rise to significant deviations from gaussianity that are not well-captured by our analytic likelihood approximation \citep[cf. Figure~11 in ][]{hennawi2024}.
As a result, our reference model is highly overconfident and we have to apply the coverage correction procedure to obtain meaningful posteriors. All inference test results for this IGM transmission model are shown in Figure~\ref{fig:coverage_sims}. Coverage probability largely remains unchanged with respect to latent dimension and red-side wavelength coverage. The effect of improving coverage for lower resolution spectra becomes more prominent than for the analytic IGM transmission model and also becomes apparent for lower signal-to-noise spectra. For spectra with $\mathrm{S}/\mathrm{N}$ as low as $2$ or $\mathrm{FWHM} \gtrsim 1000 \mathrm{km}/\mathrm{s}$, the inferred posteriors are close to passing the inference test without coverage correction. We note that such low $\mathrm{S}/\mathrm{N}$ and low resolution spectra naturally come at the price of decreasing inference precision. As shown in Appendix~\ref{app:param_scan_sims}, however, all precision gains from high $\mathrm{S}/\mathrm{N}$ or high resolution spectra are neutralized by their deteriorating coverage behavior. Rebinning spectra to coarse $\gtrsim 500 \mathrm{km}/\mathrm{s}$ velocity pixel scales (corresponding to $\mathrm{FWHM} \gtrsim 1000 \mathrm{km}/\mathrm{s}$ in our analysis) therefore provides a simple way of reducing the computational complexity of the inference problem (or the observational cost of obtaining the required spectra) without losing inference precision. 
An improved prescription of the likelihood (Eq.~(\ref{eq:likelihood}), and especially Eq.~(\ref{eq:P_t_theta})) would be required to make use of the full information encoded in the high-resolution spectra. A promising avenue for this is simulation-based inference, i.e. training a neural network-based estimator to learn the true non-Gaussian likelihood.

\section{Parameter Scan for the Full-Simulation Model}
\label{app:param_scan_sims}

\begin{figure*}
    \centering
	\includegraphics[width=0.98\textwidth]{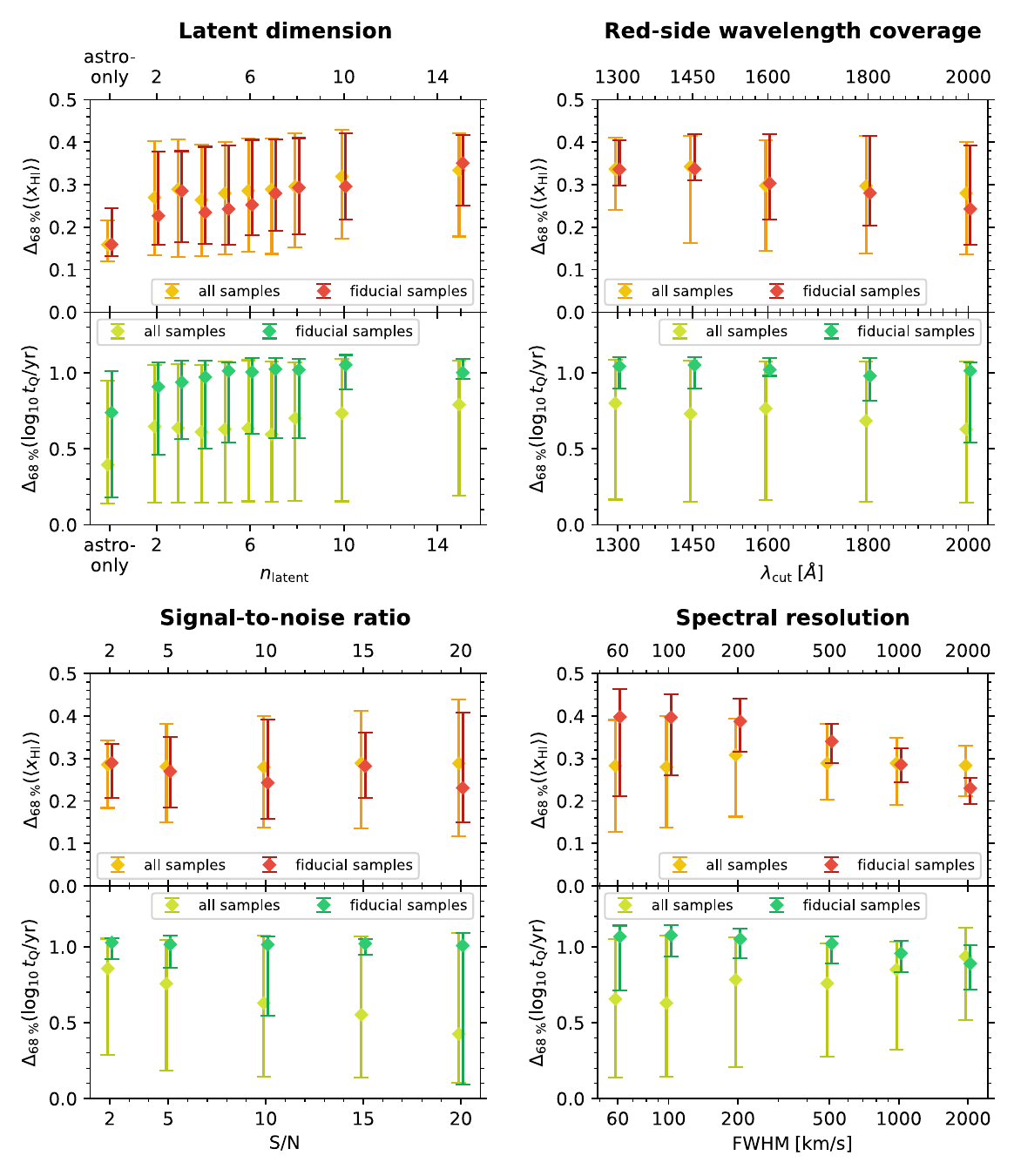}
    \caption{Like Figure~\ref{fig:mix_precision}, but for the full-simulation IGM transmission model. Trends are less pronounced due to the additional stochasticity, and the increasingly significant coverage corrections required at high $\mathrm{S}/\mathrm{N}$ or high resolution undermine all precision gains or even lead to a decrease in precision after accounting for overconfidence.}
    \label{fig:mix_precision_sims}
\end{figure*}

\begin{figure*}
    \centering
	\includegraphics[width=\textwidth]{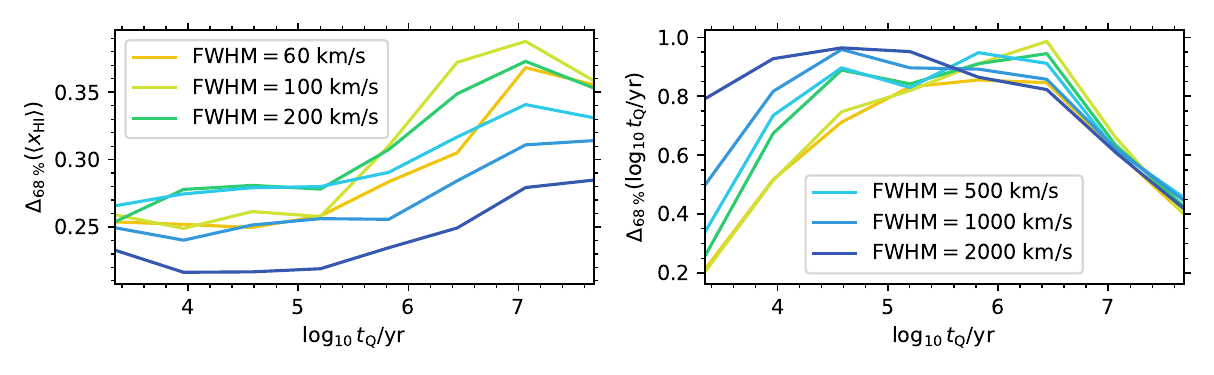}
    \caption{Inference precision with respect to IGM neutral fraction $\langle x_\mathrm{HI} \rangle$ (left) and quasar lifetime $t_\mathrm{Q}$ (right) as a function of lifetime for different choices of spectral resolution. The remaining hyper-parameters are always fixed to their reference values $n_\mathrm{latent} = 5$, $\lambda_\mathrm{cut} = 2000\,\text{\AA}$, $\mathrm{S}/\mathrm{N} = 10$. IGM transmission skewers are generated by the full-simulation model introduced in Section~\ref{sec:analytic}. Each curve shows median precision values with respect to an ensemble of $250$ mock quasars at fixed neutral fraction $\langle x_\mathrm{HI} \rangle = 0.5$, grouped into lifetime bins. All precision values are based on reweighted marginal posterior PDFs, assuring that each mock ensemble passes a marginal coverage test with respect to the two astrophysical parameters $\langle x_\mathrm{HI} \rangle$ and $t_\mathrm{Q}$. Neutral fraction precision decreases after accounting for overconfidence, independently of lifetime, whereas lifetime precision keeps improving for short-lived objects.}
    \label{fig:sims_fwhm_lifetime_prec}
\end{figure*}

We confirm in this section that the inference precision trends identified in Section~\ref{sec:param_scan} also hold when reionization is modelled realistically based on cosmological simulations. To that end, we re-ran all simulations presented in Section~\ref{sec:param_scan} with the full-simulation IGM transmission model (Section~\ref{sec:sims}) in place of the analytical one (Section~\ref{sec:sims}). Figure~\ref{fig:mix_precision_sims} summarizes all dependencies of $\langle x_\mathrm{HI} \rangle$- and $t_\mathrm{Q}$-inference precision on the four hyper-parameters $n_\mathrm{latent}$, $\lambda_\mathrm{cut}$, $\mathrm{S}/\mathrm{N}$ and $\mathrm{FWHM}$ (analogous to Figure~\ref{fig:mix_precision} for the analytical IGM transmission model). Apart from the additional stochasticity that dilutes the overall trends, we find the same qualitative behavior for latent dimension and red-side wavelength coverage. Rather than higher precision for less noisy spectra, however, we observe identical $\langle x_\mathrm{HI} \rangle$-precision values independently of the signal-to-noise ratio. The same is the case for lifetime precision of the fiducial subset of samples. Note that all precision values in Figure~\ref{fig:mix_precision} correspond to the widths of the \textit{reweighted} posterior distributions \textit{after} accounting for any degree of overconfidence which becomes increasingly relevant with increasing $\mathrm{S}/\mathrm{N}$ as seen in Appendix~\ref{app:inf_tests}: all theoretical precision gains are thus neutralized by the coverage corrections applied to the posteriors whose widths are shown in Figure~\ref{fig:mix_precision_sims}. Without these corrections, precision would improve monotonically with increasing $\mathrm{S}/\mathrm{N}$ as intuitively expected. The picture is similar for spectral resolution: after accounting for overconfidence, inference precision in $\langle x_\mathrm{HI} \rangle$ hardly changes with $\mathrm{FWHM}$, and for the fiducial set of objects even \textit{decreases} with higher resolution. Lifetime precision, on the other hand, \textit{does} improve to some extent when considering samples from the full prior range, but we find the opposite trend for the fiducial set of samples. All curves have in common that the scatter increases significantly towards higher resolution, i.e., a certain subset of spectra benefits notably from increasing resolution whereas the opposite is the case for the rest.

To investigate this behavior more carefully, we fix the value of the IGM neutral fraction to $\langle x_\mathrm{HI} \rangle = 0.5$ and consider $250$ new mock spectra, only varying the lifetime of the quasars. Figure~\ref{fig:sims_fwhm_lifetime_prec} depicts inference precision as a function of lifetime for this new set of samples. We can see that $\langle x_\mathrm{HI} \rangle$-precision approximately changes equally with resolution at all lifetime values considered, whereas the bulk of improvements in $t_\mathrm{Q}$-precision can be attributed to low-lifetime objects. The intermediate to long-lifetime range stays constant or even decreases with better resolution, in agreement with the fiducial curve in the bottom right panel of Figure~\ref{fig:mix_precision_sims}. %
This shows that our Gaussian likelihood approximation currently limits us to low resolutions because Eq.~(\ref{eq:P_t_theta}) is not able to capture the non-Gaussian information encoded in higher-resolution spectra, an obstacle that could be overcome with simulation-based inference.

\bsp	%
\label{lastpage}
\end{document}